\documentclass[aps,prb,twocolumn,showpacs,preprintnumbers,notitlepage,superscriptaddress,floats]{revtex4-2}
\usepackage{bbm}

\usepackage{tikz}

\usepackage{graphicx}
\usepackage{graphics}
\usepackage{amsmath}
\usepackage{amssymb}
\usepackage{amsfonts}
\usepackage{dsfont}
\usepackage{braket}
\usepackage{color}
\usepackage{braket,slashed}
\usepackage[mathscr]{euscript}
\definecolor{darkblue}{rgb}{0, 0, 0.8}
\usepackage[colorlinks=true, breaklinks=true, linkcolor=red, citecolor=blue, urlcolor=blue]{hyperref} 
\usepackage{subfigure}
\usepackage{xfrac}
\usepackage{bm}
\usepackage{kantlipsum}
\usepackage{enumitem}
\usepackage{tikz}
\usepackage{framed}
\usepackage{graphicx}
\usepackage{subfigure}
\usepackage{cleveref}
\usepackage{adjustbox}

\usepackage{xcolor}

\allowdisplaybreaks[1]

\newcommand{\code}[1]{\texttt{#1}}

\begin{document}

\title{Real-time dynamics of a critical Resonating Valence Bond spin liquid}

\author{Ravi Teja Ponnaganti}
\affiliation{Laboratoire de Physique Th\'{e}orique, C.N.R.S. and Universit\'{e} de Toulouse, 31062 Toulouse, France}

\author{Matthieu Mambrini}
\affiliation{Laboratoire de Physique Th\'{e}orique, C.N.R.S. and Universit\'{e} de Toulouse, 31062 Toulouse, France}

\author{Didier Poilblanc}
\affiliation{Laboratoire de Physique Th\'{e}orique, C.N.R.S. and Universit\'{e} de Toulouse, 31062 Toulouse, France}

\begin{abstract}
Implementation of the hardcore-dimer Hilbert space in cold Rydberg-atom simulators opens a new route of investigating real-time dynamics of dimer liquids under Hamiltonian quench. Here, we consider an initial Resonating Valence Bond (RVB) state on the square lattice realizing a critical Coulomb phase with algebraic and dipolar correlations. Using its representation as a special point of a broad manifold of SU($2$)-symmetric, translationally invariant, Projected Entangled Pair States (PEPS), we compute its non-equilibrium dynamics upon turning on inter-site Heisenberg interactions. We show that projecting the time-evolution onto the PEPS manifold remains accurate at small time scales. We also find that the state evolves within a PEPS sub-manifold characterized by a U($1$) gauge symmetry, suggesting that the Coulomb phase is stable under such unitary evolution.
 
\end{abstract}

\maketitle

\section{Introduction}

The search for spin liquids in condensed matter materials is a very rapidly developing area of quantum magnetism~\cite{Savary2016}. Spin liquids, as the Resonating Valence Bond (RVB) state proposed by Anderson~\cite{Anderson1973},  are prototypical states of matter showing no symmetry breaking, even down to zero temperature, due to enhanced zero-point quantum fluctuations.  Their highly entangled nature leads to unique physical aspects~\cite{Poilblanc2012,Chen2018a}, such as emerging non-local excitations, topological properties, etc...

Interestingly, spin liquids have also become accessible to other experimental set-ups based on ultracold atoms loaded into a two-dimensional optical lattice~\cite{Rutkowski2016} or in two-dimensional Rydberg arrays~\cite{Alicea2022,Semeghini2021} offering an alternative route to emulate quantum simulators~\cite{giudici2022,cheng2022}, in addition to condensed matter superconducting circuits~\cite{Satzinger2021}. 

\begin{figure}
    \centering
\includegraphics[width=\columnwidth]{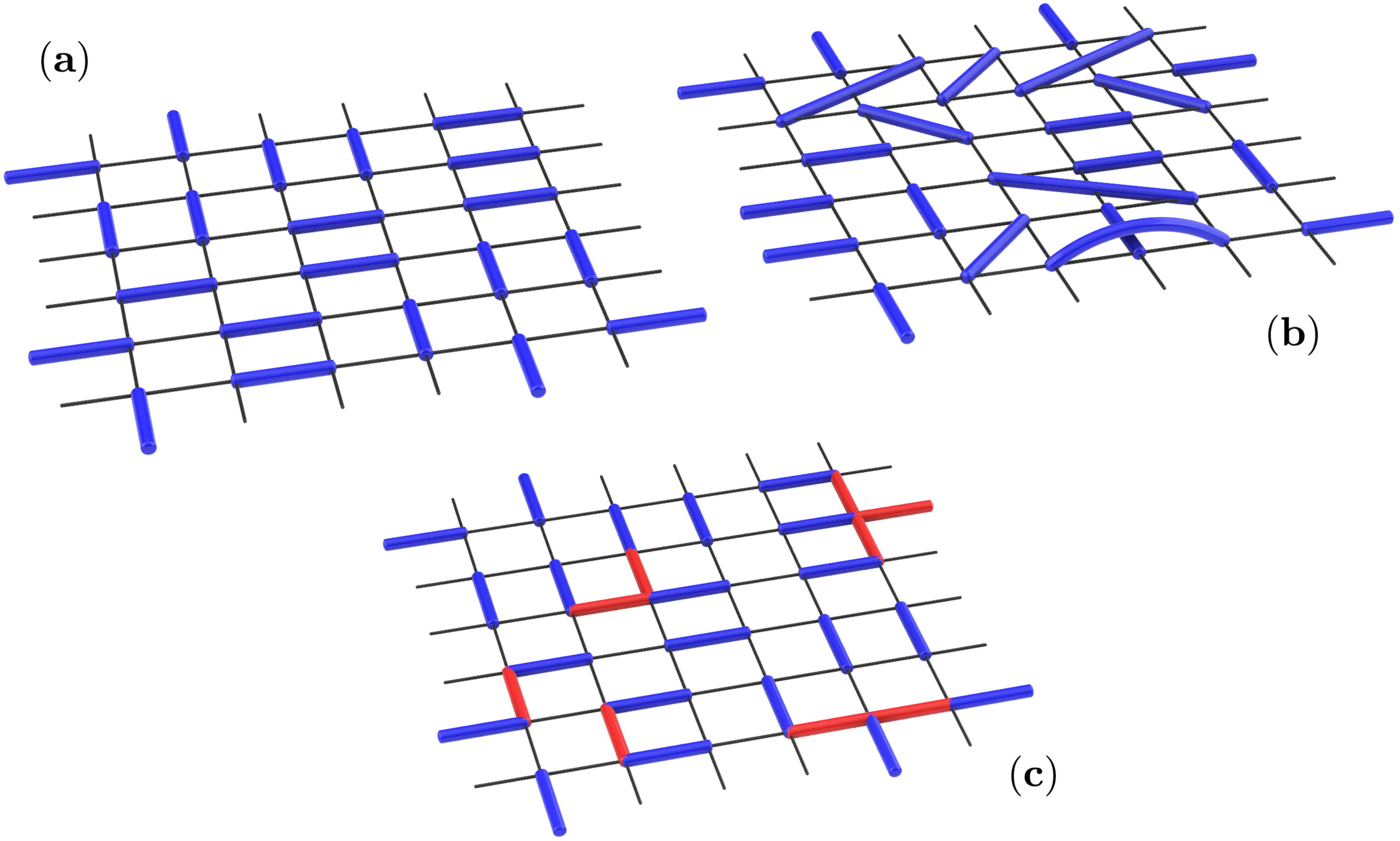}
    \caption{
  (a) A typical valence bond configuration of the NN RVB state.  
(b) A valence bond configuration including longer range singlets. (c) A typical configuration under time-evolution of the NN RVB state. Blue (red) dimers are singlet bonds build from two NN virtual spin-1/2 (spin-1) of the PEPS. The characteristic U($1$) gauge symmetry of the Coulomb phase is broken in state (b). 
    }
    \label{fig:rvb}
\end{figure}

The rapid progress in cold atom experiment setups call for new efficient theoretical tools to investigate non-equilibrium dynamics of isolated pure quantum systems. Apart from one-dimension for which efficient techniques exist~\cite{Alba2017,Robinson2021}, computing the non-equilibrium dynamics that follows a quantum quench~\cite{Das2020} is a tedious task in two-dimensional quantum spin systems~\cite{Aditi2018,Sanchez2018}. Here we shall investigate the non-equilibrium dynamics of the RVB state, the most paradigmatic example of a spin liquid. For simplicity, we shall consider the case of the square lattice. A pictorial representation of the nearest-neighbor RVB state is shown on Fig.~\ref{fig:rvb}(a), consisting only of resonating nearest-neighbor (NN) valence bond configurations~\cite{Anderson1973}. Interestingly, it has been shown that the NN RVB state exhibits critical dimer-dimer correlations connected to a local U(1) gauge symmetry and  characteristic of a Coulomb phase~\cite{Moessner2003,Moessner2011}. One of the main goals of this work is to investigate the stability of the Coulomb phase following a quantum quench as well as the dynamics of its entanglement. 

\begin{figure}
	\centering
	\includegraphics[width=0.9\columnwidth]{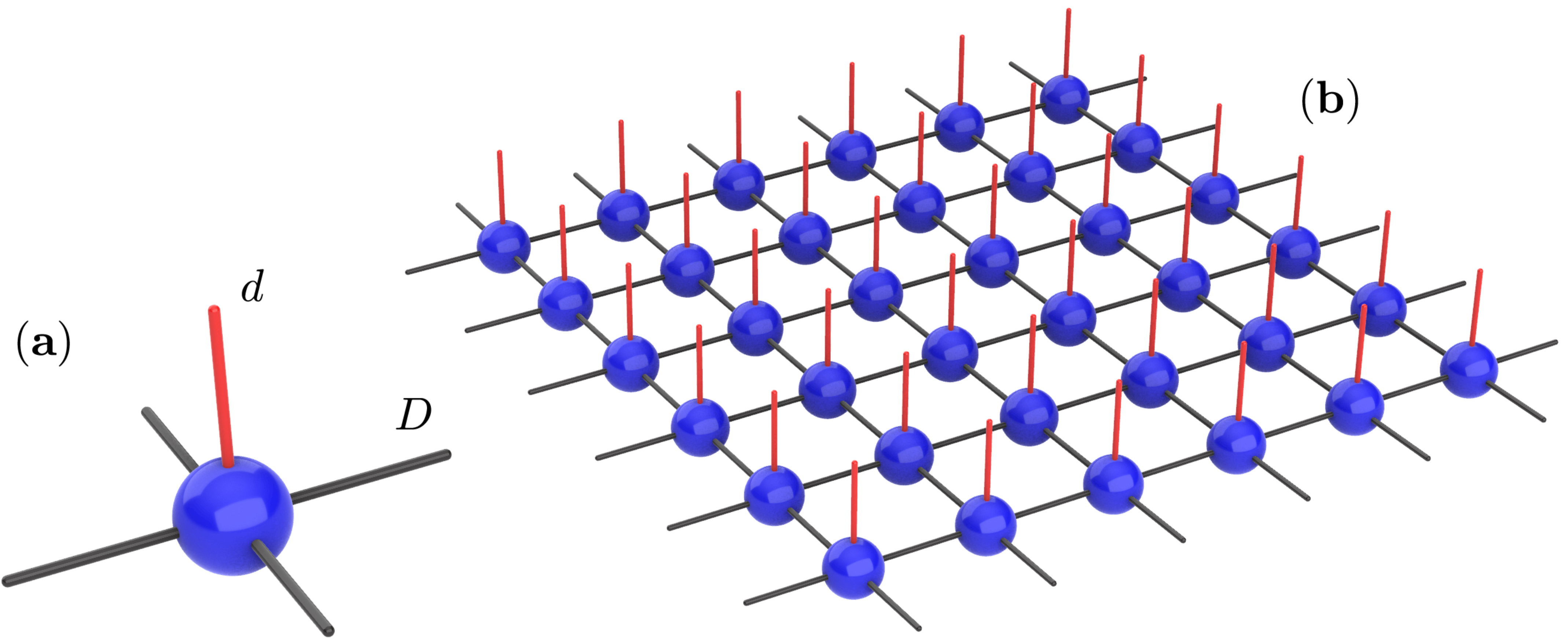}
	\caption{
		(a) The (time-dependent) site tensor (blue dot) contains 4 virtual bonds (in black) of dimension $D$ and a physical leg (in red) of dimension $d=2$ (spin-1/2 or "qubit").  
		(b) The iPEPS is obtained from an infinite square-lattice array of site tensors contracted over all virtual indices. 
	}
	\label{fig:TN}
\end{figure}

For such a purpose, through out this paper, we shall use the tensor network formalism using a variational Projected Entangled Pair State (PEPS) ansatz of the time-evolving many-body wave function~\cite{Mambrini2016}. This procedure is, in spirit, similar to ref.~\cite{giudici2022} where the preparation dynamics of a
Rydberg quantum simulator is approximated by projecting it on a tensor network
manifold. In our case, the PEPS is defined by a unique time-dependent tensor ${\cal A}(t)$ placed on all the sites of a two-dimensional square lattice. Using Penrose graphical representation~\cite{Penrose1971}, the on-site tensor and its corresponding tensor network are shown in Fig.~\ref{fig:TN}(a) and (b). 
Hence ${\cal A}(t)$ encodes locally, at all times, the coefficients of the many-body wave function $|\Psi(t)\big>$ in the exponentially-large $S_z$-basis $\{\sigma_1,\sigma_2,\cdots,\sigma_N\}$, $\sigma_i=\pm 1/2$, as shown in Fig.~\ref{fig:TN}(b). Note that the entanglement (which grows with time) is controlled by the (virtual) bond dimension $D$ of the tensor ${\cal A}(t)$. 
For simplicity we shall also take the limit of an infinite system, $N\rightarrow\infty$, using the infinite-PEPS (iPEPS) framework~\cite{Jordan2008}. 

The paper is organized as follow; First, one describes the set-up of the global quench in Section~\ref{sec:quench}: A description of the physical nature of the initial spin liquid state and a summary of the standard Trotter-Suzuki procedure used to obtain the time-evolution are given in Subsections \ref{subsec:rvb} and \ref{subsec:ts}, respectively. We detail the implementation of the simple update (SU) framework keeping track of the space-group and spin-SU($2$) symmetries of the initial state and of the Hamiltonian in Section~\ref{sec:num}. This procedure involves the determination of the optimal virtual space described in Subsection~\ref{subsec:virtual} and a careful gauge fixing implementation described in Subsection~\ref{subsec:gauge}. Details on the recovery of the exact point group and SU($2$) symmetries at each Trotter step is left to Appendix~\ref{appendix:sym}. Results are provided and discussed in Section~\ref{sec:results}: First, the SU singular value spectrum is analyzed as a function of propagating time in Subsection~\ref{subsec:sv}. Then, we propose the reversal probability (Lodschmidt echo) as an interesting tool to measure truncation errors in Subsection~\ref{subsec:fidelities}. The relevance of the finite-$D$ ansatz is controlled by investigating the energy conservation in Subsection~\ref{subsec:energy}. Finally, the critical property of the time-evolved state is investigated via the scalings of the correlation length and the entanglement entropy of the boundary state in Subsection~\ref{subsec:critical}. A summary is given and possible further developments are discussed in Section~\ref{sec:conclusion}. Additional material is provided in two other Appendices. The new algorithms specific to the factorisation of {\it complex symmetric} tensors appearing in this work are described in Appendix~\ref{appendix:factorisation}. Other specificities of the tensor contraction algorithm are provided in Appendix~\ref{appendix:CTMRG}. 

\section{Quench protocol} 
\label{sec:quench}

\subsection{Resonating Valence Bond States and PEPS representations}
\label{subsec:rvb}

Let us first start by describing the simple system setup we have considered. Our initial quantum state $|\Psi_0\big>$ is a Resonating Valence Bond (RVB) spin liquid on an infinite square lattice and a global Hamiltonian quench is assumed, at time $t=0$, by turning on the antiferromagnetic nearest-neighbor (NN) 
Heisenberg Hamiltonian,

\begin{equation}
      {\cal H}(t)= 
\begin{cases}
    \;0,                                  & \text{for } t \le 0\\
    \;H=\sum_{\langle x,y \rangle}H_{xy},   & \text{for } t > 0
\end{cases}
\end{equation}
where 
\begin{equation}
H_{xy}=J{\bf S}_x\cdot{\bf S}_y.
\end{equation}

The NN RVB state, consisting of resonating NN singlets as shown on Fig.~\ref{fig:rvb}(a), is in fact a special point of an extended one-dimensional RVB family~\cite{Chen2018a} including longer-range valence (singlet) bonds (see Fig.~\ref{fig:rvb}(b)).
This RVB family is conveniently represented by a simple PEPS manifold spanned by two single-site tensors with full lattice ($C_{4v}$) and spin-rotation (SU($2$)) symmetries; the four virtual legs have virtual space ${\cal V}_0=0\oplus\frac{1}{2}$ and are contracted, whereas the physical legs correspond to the spins in the lattice~\cite{Poilblanc2012,Schuch2012} (see Fig.~\ref{fig:TN}).
Hence, one can tune the initial state by simply varying the ratio $\lambda_2/\lambda_1$ of the coefficients of the on-site tensor,
\begin{equation}
  {\cal A}(0)=\lambda_1 {\cal A}_1 + \lambda_2 {\cal A}_2 \, , 
  \label{eq:init_tensor}
\end{equation}
where the tensor ${\cal A}_1$ defines the NN RVB state and the tensor ${\cal A}_2$ induces longer-range singlets by ``teleportation". These tensors simply differ by the occupation of the spin-$0$ and spin-$1/2$ states on the four virtual bonds, $n_{\rm occ}=\{3,1\}$ and $n_{\rm occ}=\{1,3\}$, respectively.
Recent work~\cite{Chen2018a,Dreyer2020} suggested that topological order appears whenever longer-range bonds are present (pictorially shown in Fig.~\ref{fig:rvb}(b)), i.e. $\lambda_2\ne 0$, breaking the U($1$) gauge symmetry to $\mathbb{Z}_2$.
Here, we shall take advantage of the small bond dimension $D_0=3$, and of the full symmetries of our initial state and of the Hamiltonian to study the time evolution over a small time interval. 

\subsection{Time evolution}
\label{subsec:ts}

During its time evolution the RVB state is expected to preserve its global singlet character (SU(2) rotation symmetry) and the full lattice symmetry (C$_{4v}$). 
In order to compute, for $t>0$, 
\begin{equation}
    |\Psi(t)\big>=\exp{(-iHt)}|\Psi_0\big>
\end{equation} 
we have used a simple update (SU) method~\cite{Jiang2008}  which can be implemented in a way that preserves all the symmetries under consideration. Hereafter time $t$ will be measured in unit of $1/J$. 
Here the time-evolved state is defined by a unique on-site complex PEPS tensor of bond dimension $D>D_0$ (see Fig.~\ref{fig:TN}(a)) expanded in a fully lattice-C$_{4v}$/spin-SU(2) symmetric (real) tensor basis $\{ T_a \}$, 
\begin{equation}
\label{eq:expansion}
  {\cal A}(t)=\sum_{a=1}^M \mu_a(t) \,T_a,  
\end{equation}
with $\mu_a(t)\in\mathbb{C}$. This is a simple generalization~\cite{Mambrini2016} of the symmetric PEPS construction of the initial RVB state. 
Since entanglement grows with time, it is necessary to include new virtual states, hence increasing the bond dimension and the number of elementary tensors. Later, we will show that, for the NN RVB $\lambda_2=0$ (or $\lambda_2$ small enough) it is sufficient to consider a virtual space  $V=0\oplus\frac{1}{2}\oplus 1$, i.e. a bond dimension $D=6$, to describe the time-evolution for $t\lesssim 1$ (see Table~\ref{tab:tensors}).

\begin{table}[]
    \centering
    \begin{tabular}{c|c|c|c|c|c}
    $D$   &  ${\cal V}$& C$_{s}$ & C$_{s}$/U(1)& C$_{4v}$ & C$_{4v}$/U(1) \\
    \hline
    3   & $0 \oplus \frac{1}{2}$ & 7 & 4 & 2 & 1 \\
    6   & $0 \oplus \frac{1}{2}\oplus 1$  & 41 & 30 & 11 & 8 \\
    \hline
    \end{tabular}
    \caption{Number of SU($2$)-symmetric local tensors of bond dimension $D=3$ or $D=6$, virtual space $\cal V$, C$_s$ or C$_{4v}$ point-group symmetry, and with/without U($1$) gauge symmetry (see text). The number of $C_s$ symmetric tensors gives the number of degrees of freedom available at every time sub-step. We explicitly find that the non-U($1$) symmetric tensors keep a vanishing weight under time evolution.} 
    \label{tab:tensors}
\end{table}

\begin{figure*}
    \centering
\includegraphics[width=0.8\textwidth]{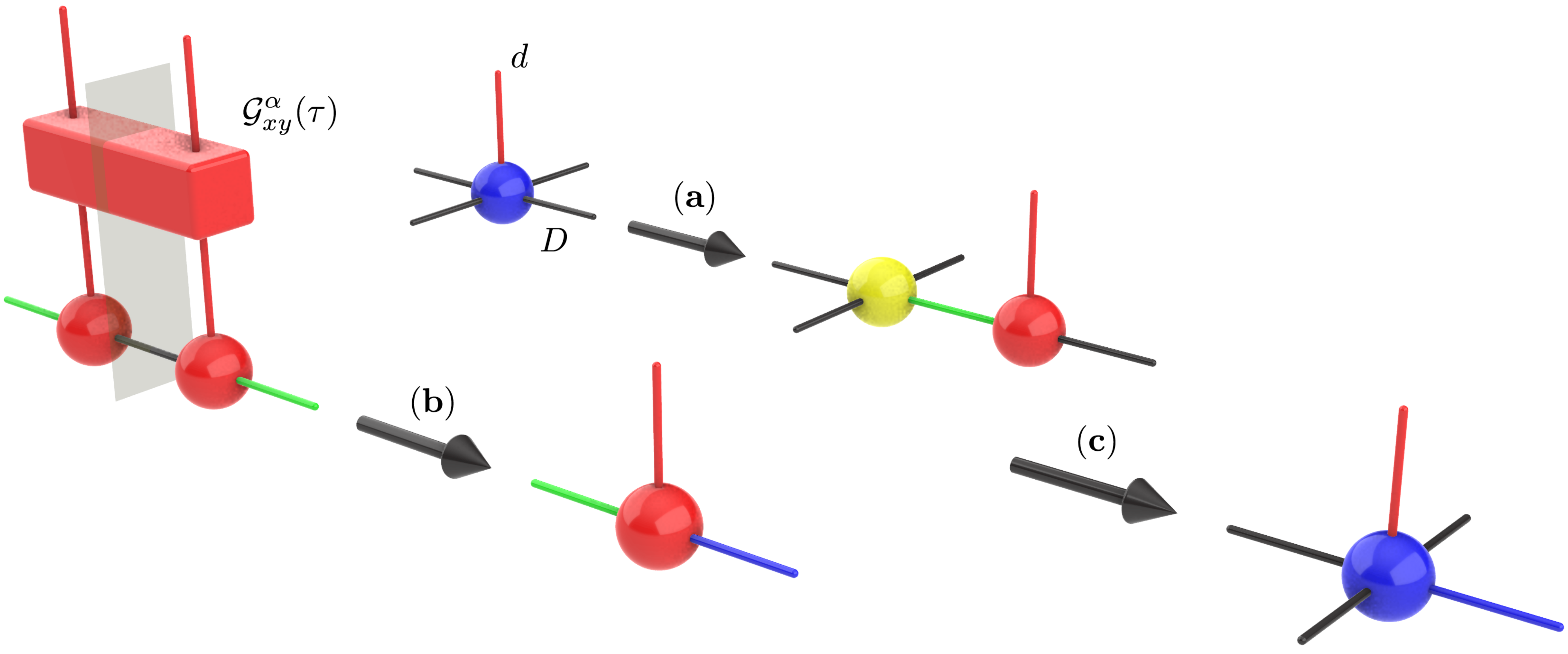}
    \caption{
  Simple update scheme: (a) the site tensor (blue sphere) is split by SVD to isolate the active bond (black leg on the red sphere); 
(b) The complex symmetric gate ${\cal G}_{xy}^{\alpha} (\tau)$ applied to the two-site bond (referred in the text as ``SU matrix'') is split using a Autonne-Takagi factorization~\cite{Autonne1915,Takagi1924} (further details are given in Appendix~\ref{appendix:factorisation}); (c) the two sides are used to reconstruct the new site tensors (blue sphere with updated blue leg). Note that for step (a) the SVD spectrum $S$ (diagonal matrix) is absorbed entirely in the active bond tensor (depicted in red), i.e. $U S V^\dagger = U  \left ( S V^\dagger \right )$ while for step (b) the Autonne-Takagi spectrum $S$ is symmetrically absorbed on both sides to fulfill the reflexion symmetry (depicted as a gray plane), i.e. $ US U^T = \left ( U \sqrt{S} \right) \left ( U \sqrt{S} \right)^T$.
    }
    \label{fig:su}
\end{figure*} 
Once the virtual space is defined, in order to obtain the time-evolution of the coefficients $\{\mu_a(t)\}$ we used a standard Trotter-Suzuki (TS) decomposition~\cite{SUZUKI1990} of the unitary time-evolution operator $$\exp{(-iHt)}=\prod_1^{N_\tau} \exp{(-iH\tau)},$$ where $\tau=t/N_\tau$ is a small time step (such that $\tau\ll 1$). The Heisenberg Hamiltonian is then split into four parts, 
\begin{equation}
\label{eq:ham_splitting}
    H=H^A+H^B+H^C+H^D\, ,
\end{equation}
each acting on one of the four staggered configurations ${\cal C}^\alpha$ of {\it disconnected} horizontal or vertical bonds labelled by $\alpha=A,B,C,D$. 
The action of the elementary time-evolution operator $\exp{(-iH\tau)}$ can then be approximated by the successive actions of four unitary gates,
\begin{eqnarray}
\label{eq:gate_product}
    {\cal G}^\alpha(\tau) &=&\exp{(-iH^\alpha\tau)} \nonumber  \\
     &=&\prod_{\rm \langle x,y\rangle \in C_\alpha}\exp{(-iH_{xy}^\alpha\tau)},
\end{eqnarray} 
involving the standard systematic TS error vanishing in the limit $\tau\rightarrow 0$.
Depending on the method, the update of the coefficients $\mu_a(t+\tau)$ under the action of all disconnected gates 
\begin{equation}
\label{eq:two_site_gate}
    {\cal G}_{xy}^\alpha(\tau)=\exp{(-iH_{xy}^\alpha\tau)}
\end{equation}
may be obtained locally (SU method) or take into account the environment around each of the disconnected $(x,y)$ bonds. 
We shall here focus on the  simple update method, which we describe below, and will report on a  time dependent variational optimization method (involving the environment) elsewhere.
Note that, at every sub-step defined by the action of all disconnected gates $\alpha$, the lattice $C_{4v}$ point-group symmetry is broken down to $C_s$ involving only the reflection w.r.t. the direction of the bonds $(x,y)$. Therefore, the updated one-site tensor has a basis decomposition (~\ref{eq:expansion}) involving a larger set of $C_s$-symmetric tensors (see Table~\ref{tab:tensors}). Only after a full step of four sub-steps $\alpha=A,B,C,D$ is the point-group $C_{4v}$ symmetry (approximately) restored.

\section{Simple update numerical algorithm}
\label{sec:num}

\subsection{Algorithmic steps}
\label{subsec:su}

The main lines of the Simple Update (SU) scheme we use can be summarized in the following steps :
\begin{enumerate}
    \item The tensors at sites $x$ and $y$ are first split by singular value decomposition (SVD) to isolate the active bond on which the complex symmetric gate ${\cal G}_{xy}^\alpha(\tau)$ is applied (see Fig.~\ref{fig:su}(a)).
    \item The resulting two-site symmetric complex matrix (``SU matrix" of Fig.~\ref{fig:su}(b)) is decomposed using an Autonne-Takagi factorization~\cite{Autonne1915,Takagi1924} and Appendix~\ref{appendix:factorisation} for details.
    \item The two sides are used to reconstruct the first update of the tensors at site $x$ and $y$ shown in Fig.~\ref{fig:su}(c).
    \item This procedure is then repeated for the other three bonds connected to site $x$.
    \item At last, gauge fixing and projection of ${\cal A}(t+\tau)$ onto the symmetric basis $\{T_a\}$ enables us to obtain the new set of coefficients $\mu_a(t+\tau)$.
\end{enumerate}

However, the actual implementation of this method (and especially point 5.) deserves a particular attention in the context of SU(2) invariant tensors. More specifically, we stress two important issues in the course of SU : (i) Re-defining an identical gauge convention between the four bonds after the sequence of independent factorizations  and (ii) Enforcing a fully $C_{4v}$ and SU(2) evolution at each time step. These points are discussed in detail in section~\ref{subsec:gauge}.

\subsection{Determination of the relevant virtual space}
\label{subsec:virtual}

\begin{figure}[htb]
    \centering
\includegraphics[width=\columnwidth]{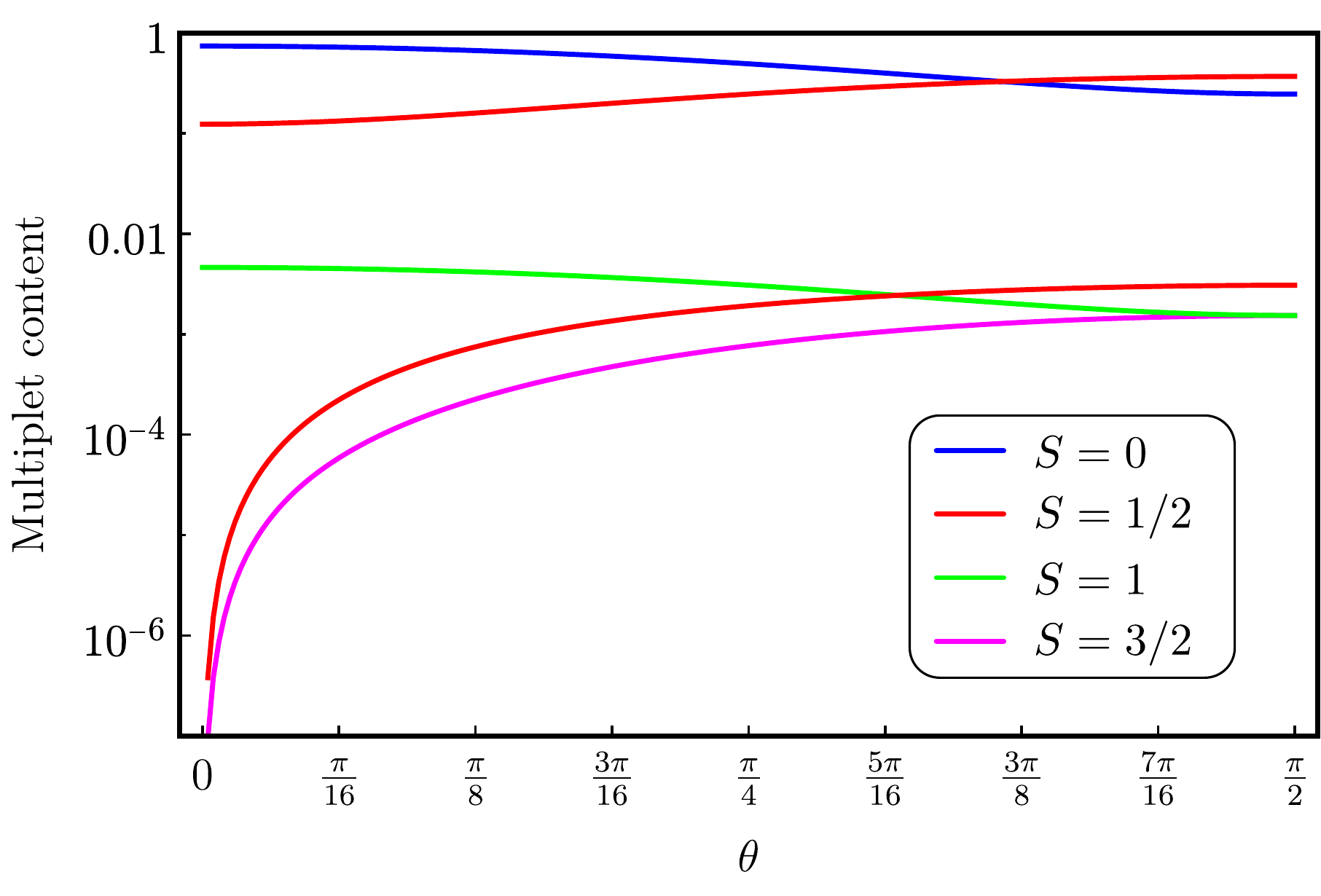}
    \caption{
  Singular values of the  Autonne-Takagi factorization of the gate of Fig.~\protect\ref{fig:su}(b) computed for $\tau=0.025$ as a function of $\theta$ with $\lambda_1=\cos (\theta)$ and  $\lambda_2=\sin (\theta)$.
    }
    \label{fig:svd_init}
\end{figure}

It is particularly interesting to first have a close look at the first application of the ${\cal G}_{xy}^A(\tau)$ gate onto the initial (general) RVB state. The result of this action will guide us to select the relevant choice of the virtual space needed to approximate the time evolution. Fig.~\ref{fig:svd_init} shows the 12 singular values of the Autonne-takagi factorization of the matrix in Fig.~\ref{fig:su}(b), grouped in SU(2) multiplets, as a function of $\theta=\tan^{-1}{(|\lambda_2|/\lambda_1)}$. Remarkably, at $\lambda_2=0$ only 6 singular values are non-zero corresponding to spin-$0$, spin-$1/2$ and spin-$1$ multiplets. This provides support for using the virtual space ${\cal V}=0\oplus\frac{1}{2}\oplus 1$ to describe time evolution at finite time $t$. Such a PEPS can be pictorially represented in the RVB language as in Fig.~\ref{fig:rvb}(c). The range of validity in time of this approximation will be discussed later on. Note that, when turning on $\lambda_2$ the additional spin-$\frac{1}{2}$ and spin-$\frac{3}{2}$ multiplets acquire some weights which increase with increasing $\lambda_2$ and, hence, could not be neglected anymore. 

\begin{figure*}
    \centering
\includegraphics[width=0.86\textwidth]{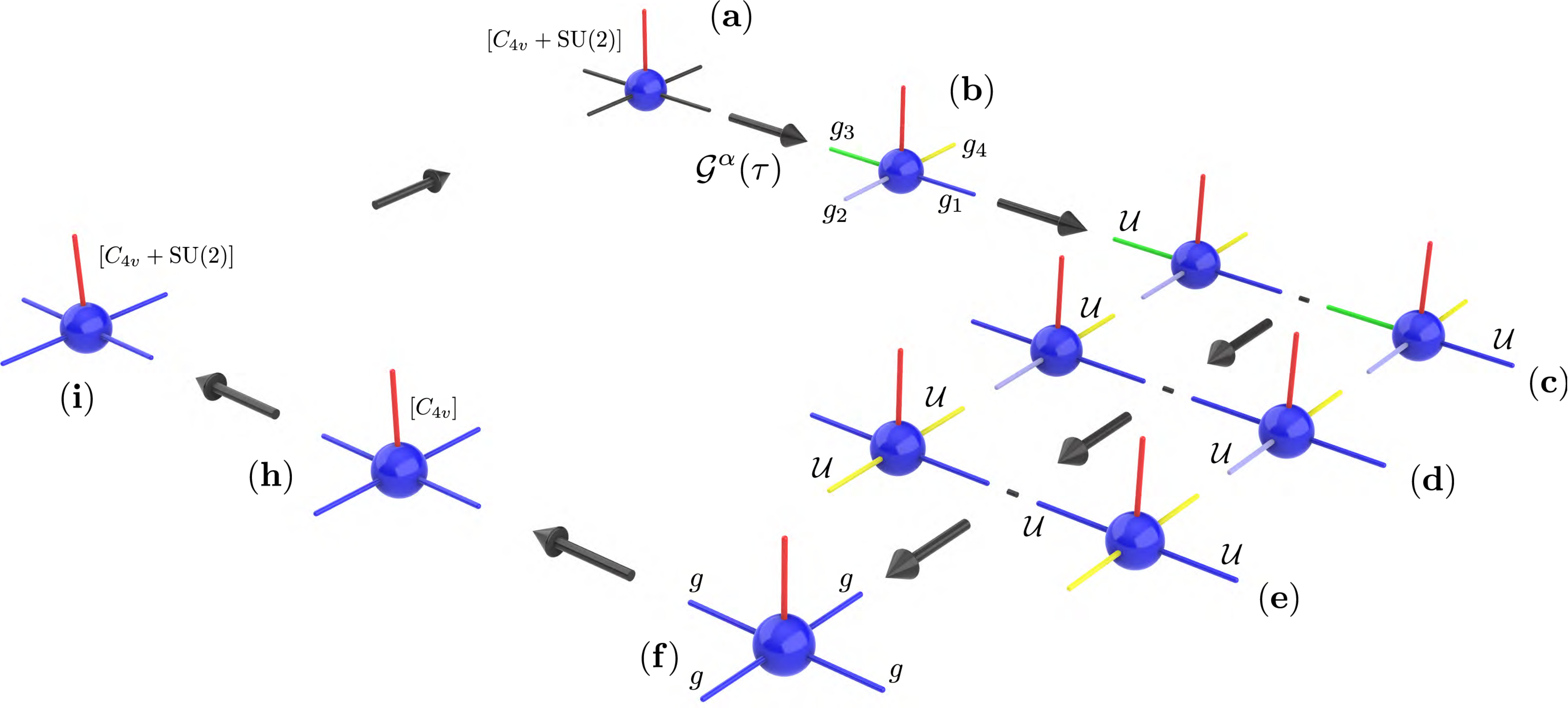}
    \caption{Gauge fixing and symmetries in SU : Starting from a fully SU(2) and $C_{4v}$ symmetric site tensor (a) the four legs are updated using the SU method (see Fig.~\ref{fig:su}) using ${\cal G}^{\alpha} (\tau)$ gate. (b) Due to gauge freedom the basis used is generically different on the four virtual legs (depicted as distinct colors). (c-e) A uniform fixed gauge is obtained in a 3-step process in which the norm of the 2 reflection dissymetries (c-d) and rotation dissymetry (e) are minimized by a complete enumeration of all possible basis (see Eq.(\ref{eq:gauge_fix})). The resulting tensor (f) is expressed in a uniform gauge and is approximately $C_{4v}$ symmetric (up to $\tau^2$ corrections). (h) The exact $C_{4v}$ symmetry is explicitly restored. (i) Projection into the SU(2) symmetric tensor basis leads to a SU(2) and $C_{4v}$ symmetric tensor suitable for the next $\tau$-step evolution.
    }
    \label{fig:su_details}
\end{figure*}

\subsection{Gauge fixing} 
\label{subsec:gauge}
In this Subsection we describe the implementation of the tensor symmetrisation at every time step (algorithmic step 5. mentioned in \ref{subsec:su}). Readers interested mostly by the physical problem may jump directly to Section~\ref{sec:results}.

According to Trotter-Suzuki decomposition (see Eqs.~(\ref{eq:ham_splitting}),(\ref{eq:gate_product}),(\ref{eq:two_site_gate}) and Fig.~\ref{fig:su}), starting from a fully symmetric tensor (Fig.~\ref{fig:su_details}(a)), four Autonne-Takagi transformations are required at every time step. Since tensors are SU(2) invariant, the singular values spectra splits into degenerate sectors corresponding to the various multiplets (see e.g. Fig.~\ref{fig:svd_init}). As a direct consequence, the unitary transformation involved in the factorization is {\em not unique}. More precisely, any block SU(2) rotation acting on each multiplet subspace ($S=0, S=1/2, \ldots$) leads to an equally  valid decomposition. This results in a continuous gauge freedom and generically a mismatch of the basis used on the four virtual legs (see Fig.~\ref{fig:su_details}(b)), which prevents any further computation.

If the gauge is uniformly fixed on the four virtual legs, the updated tensor remains symmetric under $C_{4v}$ transformations up to $\tau^2$ corrections, which corresponds to the lowest order where non-commutativity effects occurs in the Trotter-Suzuki decomposition. This fact leads to a natural criterion to fix a uniform gauge on all four virtual legs, optimizing the tensor point group symmetry. However minimizing the tensor dissymetry under a continuous set of unitary transformations reveals intractable in practice.

A way to circumvent this problem is to apply to the tensor a {\em multiplicative} random noise (controlled by its amplitude $\varepsilon$) before performing the 4-step update. This kind of noise breaks SU(2) symmetry but preserve the U(1) symmetry related to charge conservation ($S_z$ in the context of SU(2)). Hence the singular values spectra appearing in the various decompositions get the structure of slightly splited SU(2) multiplets. Interestingly, in this scheme, $\varepsilon$ determines the {\em relative} splitting of the noise-free multiplet singular value. This makes it possible to clearly identify the multiplets in the spectra, regardless of the amplitude of the considered singular value.

But, even more importantly, disorder reduces drastically the gauge freedom to a {\em discrete} set of transformations. Keeping in mind the properties of  Autonne-Takagi transformation detailed in Appendix \ref{appendix:takagi} (in particular eq.~\ref{eq:takagi_gauge}), the gauge transformation relating two equivalent Autonne-Takagi factorizations {\em in the case of a non-degenerate spectrum} is just a diagonal matrix with $\pm 1$ entries. On top of this, the multiplicative noise can cause a reordering of states inside each multiplet in a way that does not match the canonical order (e.g $S_z=+1,0,-1$ for $S=1$). Hence the most general gauge transformation summarizes as :
\begin{equation}
    \label{eq:gauge_fix}
{\cal U} = \begin{pmatrix}\pm 1 & & \\ & \ddots & \\ & & \pm 1\end{pmatrix} \begin{pmatrix}{\cal P}_{v_1} & & \\ & \ddots & \\ & & {\cal P}_{v_n}\end{pmatrix},
\end{equation}
where ${\cal V}=\oplus_{i=1}^n v_i$ and ${\cal P}_{v_i}$ is a permutation matrix in the $v_i$ subspace ($n$ stands here for the total number of species). Hence the maximal number of distinct transformation is 
$2^D \prod_{i=1}^{n} {\text{Dim} (v_i)} !$ which is small enough in practical applications to allow direct enumeration in the course of tensor dissymetry minimization. The dissymetry over the two reflections and $\pi/2$-rotation is described if Fig.~\ref{fig:su_details} (c-e).

\begin{figure*}
	\centering
	\includegraphics[width=0.9\textwidth]{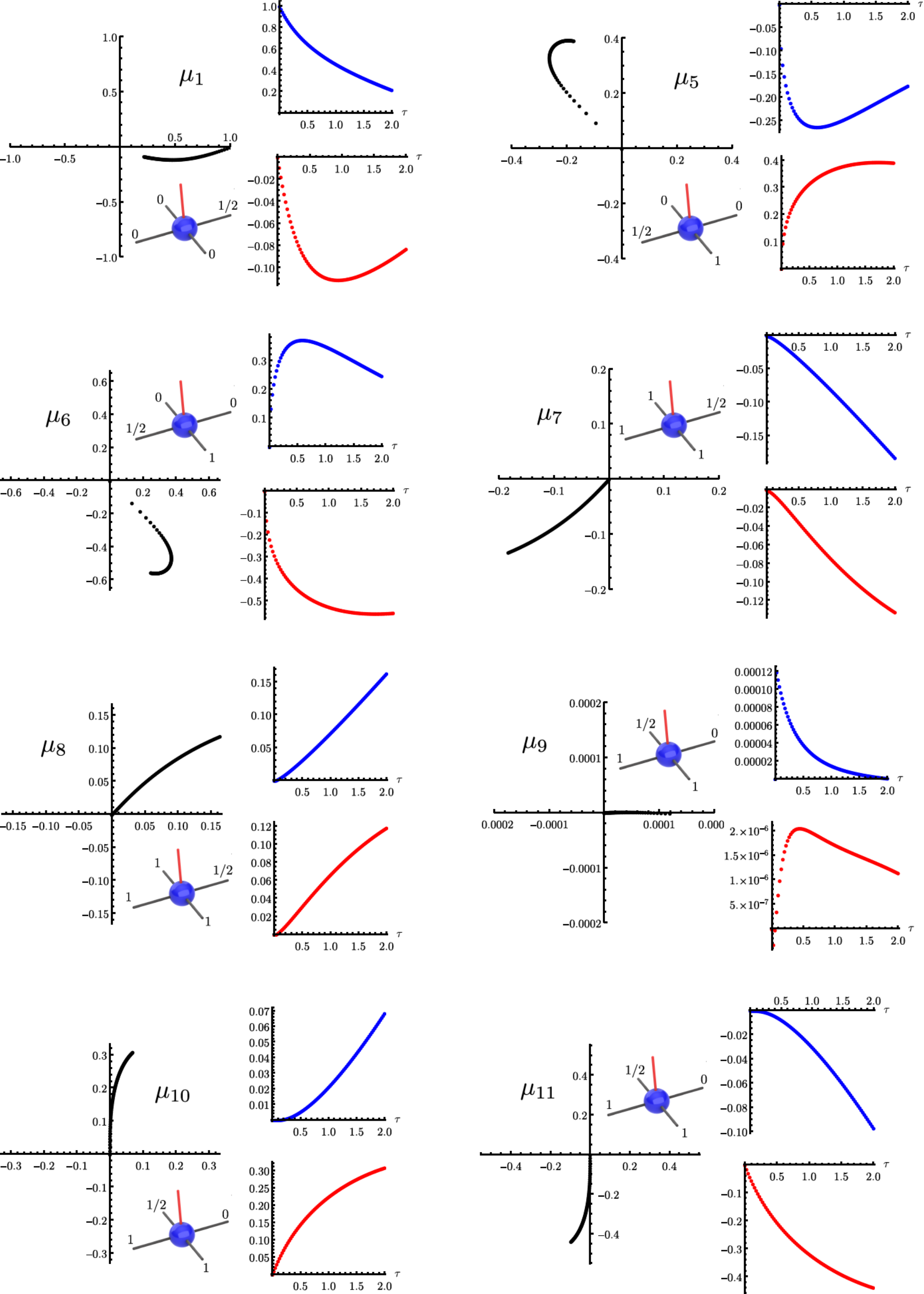}
	\caption{
		Time evolution of ${\cal A}(t)$ complex components $\mu_i$ (see Eq.~\ref{eq:expansion}) using $\tau = 0.025$ in the time range $t\in [0,2]$, starting from the initial tensor ${\cal A}(0)={\cal A}_1$ (NN RVB state). In each panel $\mu$ is displayed in the complex plane (black dots) as well as its real part (blue spheres) and imaginary part (red spheres). The components $\mu_2, \mu_3$ and $\mu_4$ of the non-U($1$) symmetric tensors are found to be identically zero and, hence, are not displayed. 
	}
	\label{fig:coeff_evol}
\end{figure*}

Once a uniform consistent gauge is fixed for the four virtual legs, the next task is to project the evolved tensor into the symmetric basis $\{T_a\}$. In this process the resulting tensor becomes fully symmetric under $C_{4v}$ and SU(2) (see Fig.~\ref{fig:su_details} (h-i)).
We  demonstrate in Appendix~\ref{appendix:sym} that, for small enough $\tau$, the modification caused to the tensor by the symmetrization is small compared to the increment of the tensor due to the time evolution itself.  Fig.~\ref{fig:coeff_evol} summarizes the time evolution of the complex components $\mu_a(t)$ of ${\cal A}(t)$ expressed in the symmetric basis, for $t\in [0,2]$, starting from the NN RVB state.

\section{Results}

In this section we shall focus on the NN RVB ($\lambda_2=0$) as an initial state for which short-time evolution with the $D=6$ PEPS ansatz has been justified above. 
\label{sec:results}

\subsection{Singular value spectrum}
\label{subsec:sv}

\begin{figure}[htb]
    \centering
\includegraphics[width=0.95\columnwidth]{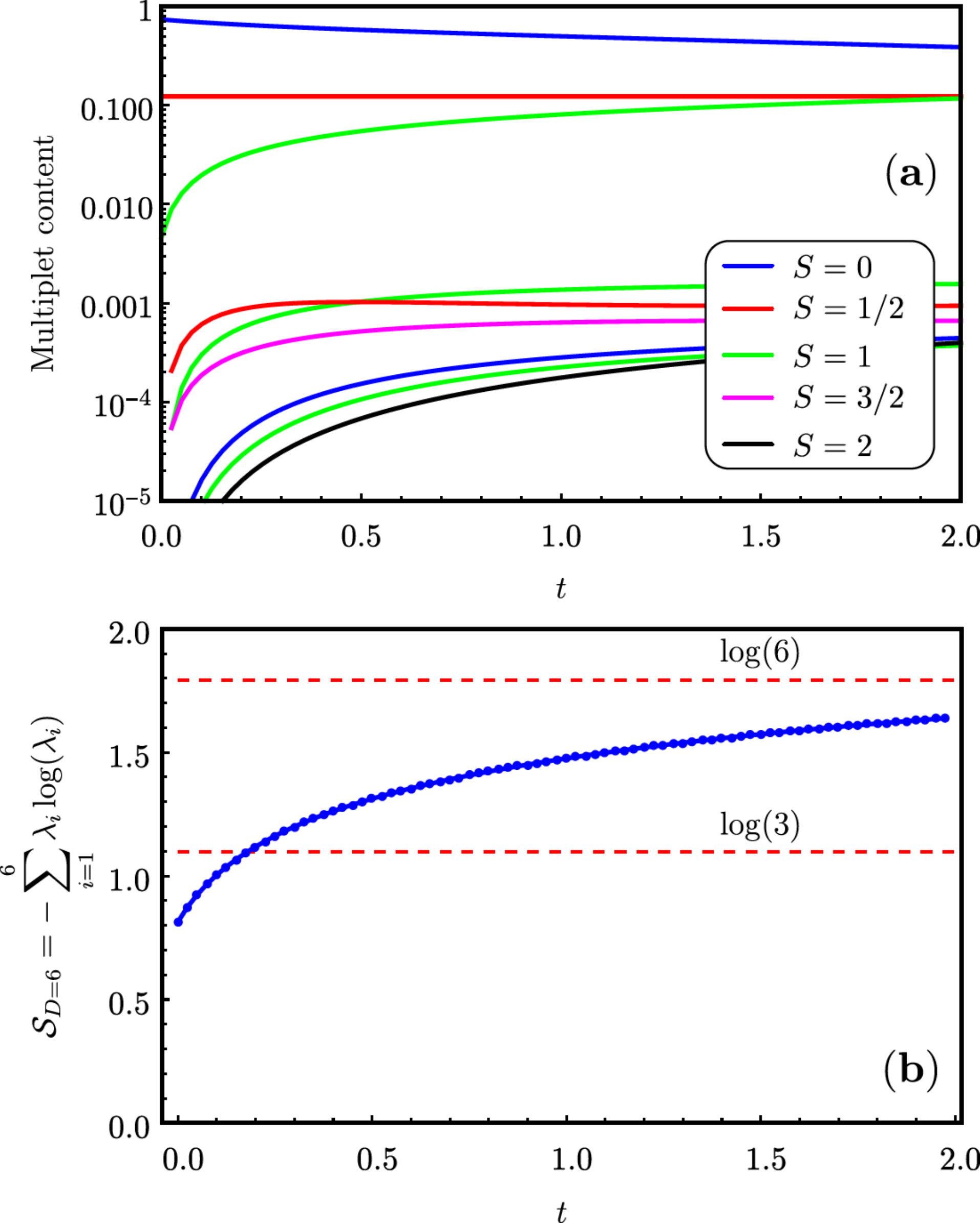}
    \caption{
  (a) Singular values $\lambda_i$ of the  Autonne-Takagi factorization of the SU gate matrix represented in Fig.~\protect\ref{fig:su}(b) and computed for $\tau=0.025$, as a function of time.
     (b) Entropy per bond as a function of time captured by the representations $0 \oplus 1/2 \oplus 1$ used in the SU procedure.
    }
    \label{fig:SU_vsTime}
 \end{figure}

Since entanglement quickly grows with time, we expect that more virtual states (i.e. larger $D$) may become necessary as time goes on. To control the validity of our fixed $D=6$ approximation, we have examined, at each step after applying the first gate ${\cal G}_{xy}^A(\tau)$ on $|\Psi(t)\big>$, the singular value spectrum of the Autonne-Takagi factorization of the SU matrix (defined in Fig.~\ref{fig:su}(b)), as a function of time $t$. We see in Fig.~\ref{fig:SU_vsTime}(a) that the 3 multiplets of largest weights always stay well separated from the rest of the spectrum. As expected, their weights tend to become equal, in order to saturate the maximum available entanglement entropy $\ln{D}=\ln{6}$ per site, as shown in Fig.~\ref{fig:SU_vsTime}(b). This is obviously an artefact of the truncation into the $D=6$ virtual space which, although involves only a small error at every step, leads to a significant cumulative error when $t\sim 1$. Hence further tests are needed to establish the maximum range of validity of our approximation.

\subsection{Fidelities and Loschmidt echo}
\label{subsec:fidelities}

\begin{figure}[htb]
    \centering
    \includegraphics[width=\columnwidth]{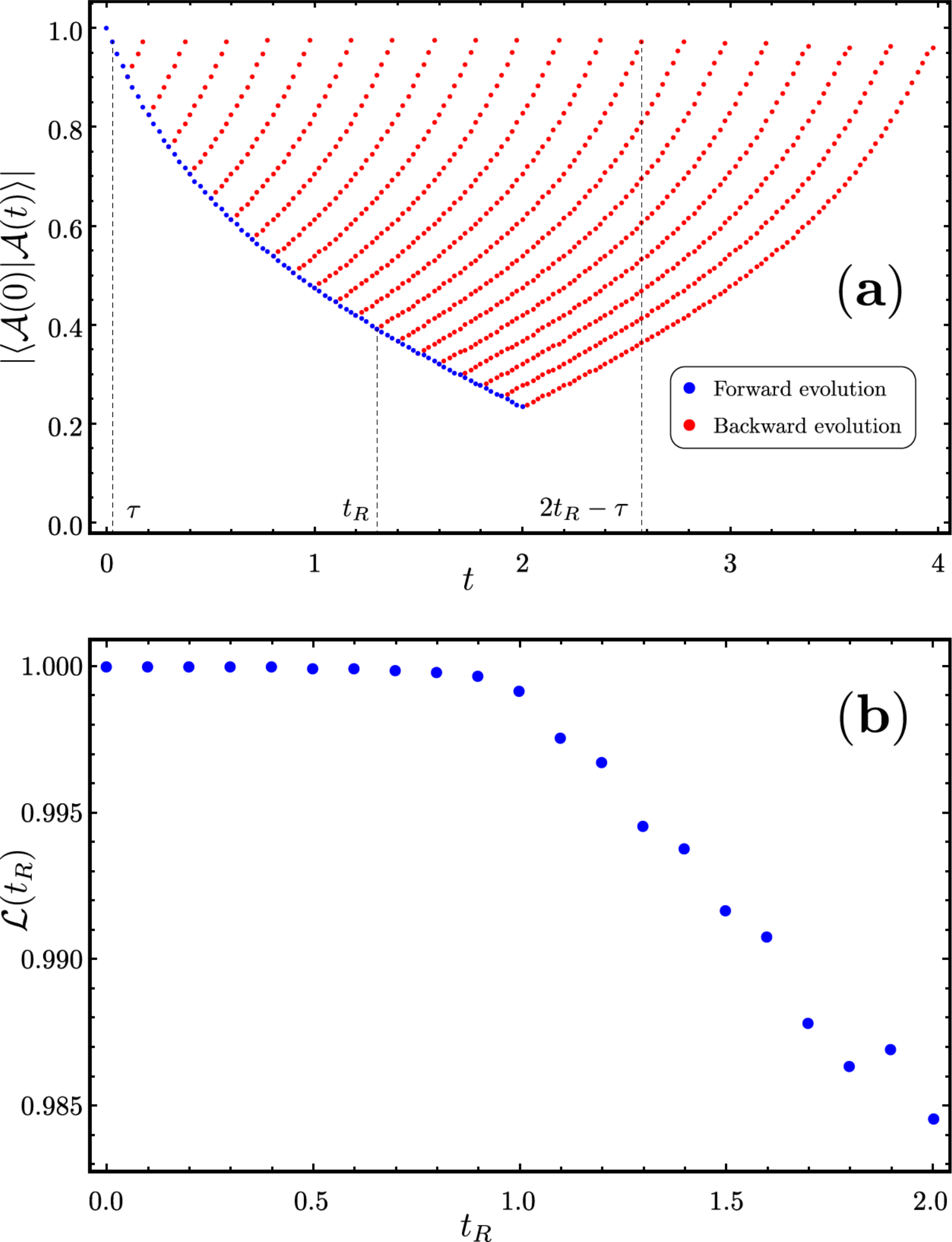}
    \caption{
    (a) Overlap $|\big<{\cal A}(t)| {\cal A}(0)\big>|$ of the time-dependent tensor with the initial tensor defining the NN RVB state ($\lambda_2=0$) as a function of time (blue dots). Red dots are used after time reversal at $t=t_R$, going backwards in time. (b) Loschmidt echo ${\cal L}(t_R) = |\big<{\cal A}(t=2t_R-\tau)| {\cal A}(\tau)\big>|$ vs $t_R$. Note that for convenience ${\cal L}(t_R)$ is not evaluated between times $0$ and $2t_R$ but between times $\tau$ and $2t_R-\tau$ to avoid dimensional jump from $D=3$ at time $0$ to $D=6$ at time $t>0$.  
    In (a) and (b) the SU method is used with $\tau=0.025$.  
    \label{fig:echo}}
 \end{figure}

In order to establish the ultimate maximum time above which our procedure breaks down, we have performed the following ``time reversal" procedure; (i) time evolution is performed with $\exp{(-iHt)}$ from $t=0^+$ to $t=t_R$; (ii) then, at $t=t_R$ time is ``reversed", $t\rightarrow -t$, i.e. time evolution is performed with $\exp{(iHt)}$. In the case of an exact unitary evolution, one should recover exactly the initial NN RVB state at time $t=t_R$. In other words, (the modulus of the overlap) $|\big<{\cal A}(t)| {\cal A}(0)\big>|$ of the time-dependent tensor with the initial tensor (defining the $\lambda_2=0$ NN RVB state) should come back to 1 at $t=2t_R$, see Fig.~\ref{fig:echo}(a). Upon increasing $t_R$, due to various cumulative errors, the evolution will stop being perfectly unitary after some time $t<t_R$. If this happens then the Loschmidt echo $|\big<{\cal A}(2t_R)| {\cal A}(0)\big>|$ will start deviating from 1. This is exactly what is observed on Fig.~\ref{fig:echo}(b) suggesting that our procedure breaks down above some ``upper bound'' time scale $t_{\rm max} \simeq 1$. Several sources of error, at different degrees, are responsible for this breakdown, including the Trotter-Susuki decomposition, the SU scheme instead of a full update scheme, the symmetrisation procedure and the restriction of a limited virtual space ($D=6$ here).

\subsection{Energy conservation}
\label{subsec:energy}

\begin{figure}[htb]
    \centering
   \includegraphics[width=\columnwidth]{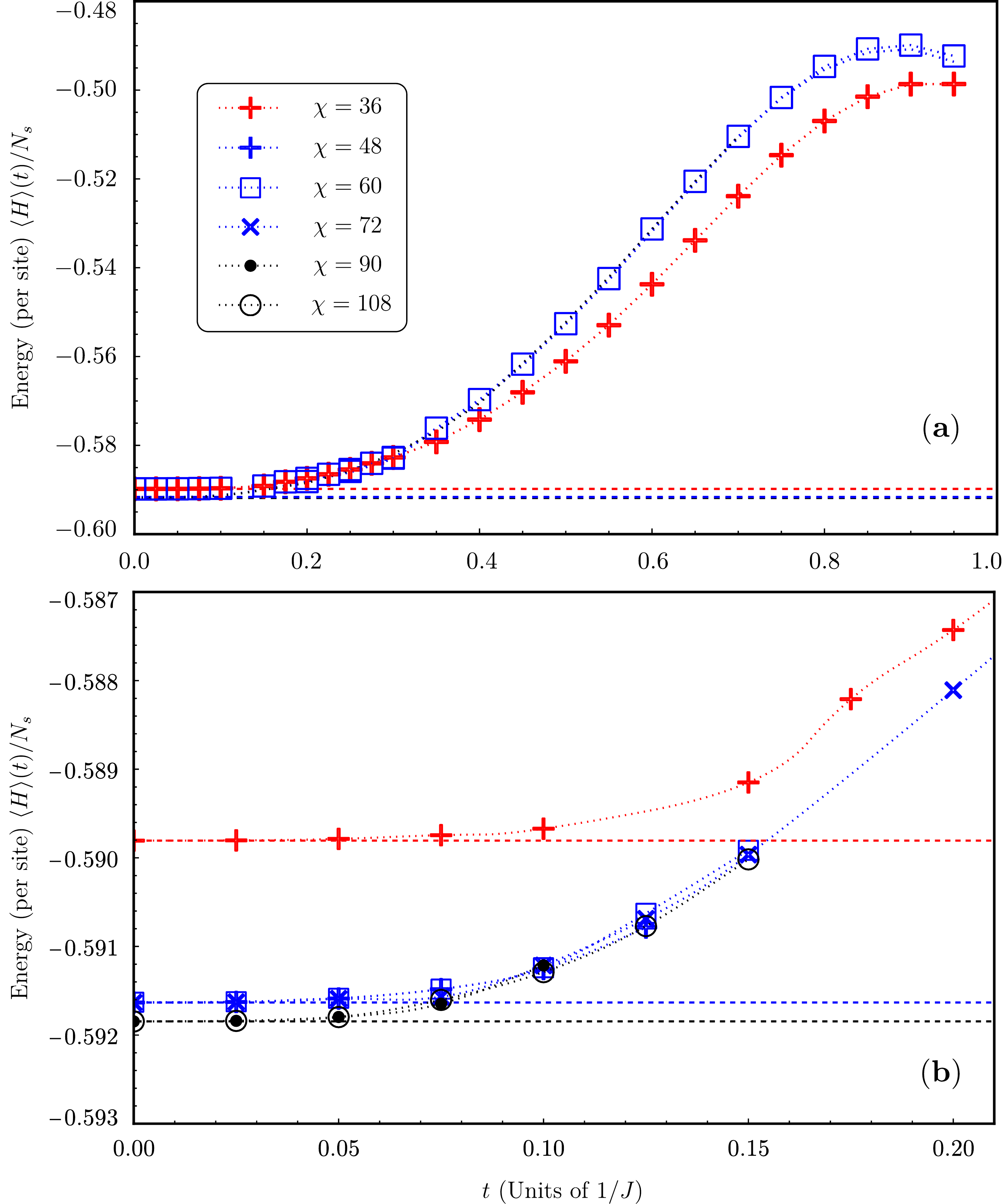}
    \caption{
     Energy per site -- ideally a constant of motion (horizontal dashed lines) -- versus time $t$, using a $D=6$ SU($2$)-symmetric iPEPS ansatz for $\chi$ varying from $D^2$
    to $3D^2$. The local tensor ${\cal A}$(t) is obtained via a SU procedure and $\tau=0.005$. (a) Full time range. (b) Zoom of the small time region.
    }
    \label{fig:energy}
\end{figure}

Under unitary time evolution, the energy -- defined as the expectation value of $H$ in the time-evolving state -- should be conserved. However, even if the simple update (symmetric) scheme succeeds (up to $t_{\rm max}$) to optimize properly the local tensor, the $D=6$ PEPS may stop accounting properly for the increase of entanglement above some intermediate time $t_{\rm max}^{\rm D=6}< t_{\rm max}$. We believe a good criterion to estimate this intermediate time scale is to examine the possible deviation of the energy w.r.t. its $t=0$ value. The latter is then computed using the SU tensors via a Corner Transfer Matrix Renormalization Algorithm (CTMRG) iPEPS algorithm~\cite{Nishino1996,NISHINO199669,Orus2009, Orus2012}. Note that the energy of the NN RVB converges slowly with the CTMRG environment dimension $\chi$ (due to its critical nature) so that several values of $\chi$ will be considered here.  Fig.~\ref{fig:energy} shows the energy per site as a function of time $t$ for $\chi$ ranging from $D^2$ to $3D^2$. At the smallest $\chi=D^2$ value, one clearly sees a plateau at small time $t\lesssim 0.25$ and, then, a clear deviation from the initial $t=0$ energy. Upon increasing $\chi$ the deviation seems to occur a bit sooner. We believe this is a sign that the restriction to $D=6$ is no longer accurate when $t>t_{\rm max}^{\rm D=6}\simeq 0.25$ and/or a full update scheme would become necessary. Notice however the energy scale used in Fig.~\ref{fig:energy}(b), showing that the energy deviation still remains relatively small up to intermediate time, e.g. $0.6\%$ at $t=0.2$.

\subsection{Critical behavior, maximum correlation length and central charge}
\label{subsec:critical}

{\it U(1) gauge symmetry --}
The basis tensors (see Table \ref{tab:tensors}) involved in the construction of our local tensor in Eq.~(\ref{eq:expansion}) can be grouped according to their occupation number of each of the virtual states on the four bonds $n_{\rm occ}=\{n_0,n_{1/2},n_{1}\}$ (with $\sum_\alpha n_\alpha=4)$. For example, the initial NN RVB state has $n_{\rm occ}=\{3,1,0\}$. The integer $n_{1/2}$ defines a U($1$) quantum number characterizing a U($1$) gauge symmetry. Because of the SU($2$) fusion rules, $n_{1/2}$ can be either $1$ or $3$ defining two separate classes of U($1$)-symmetric PEPS. We argue here that time evolution preserves the $n_{1/2}=1$ U(1) quantum number because the action of a two-site gate can only change the spin quantum number of the virtual state by 1 on the bond it is applied to. We have checked numerically that this conjecture is correct, more precisely the coefficients of all the tensors with $n_{1/2}=3$ remains identically zero, at all times. The preserved U(1) symmetry suggests that the time evolved state remains critical, as a mapping to a height representation~\cite{Moessner2003,Moessner2011} would imply.

We believe the U(1)-symmetry of the time-evolving spin liquid is not fine-tuned in the sense that it is not restricted to the $D=6$ ansatz we are using. Indeed, for any family of (singlet) SU(2)-symmetric PEPS with bond dimension $D$ one can always isolate a large subset of PEPS with U(1)-gauge symmetry (which should exhibit algebraic correlations). The (simple) argument is as follows: each local tensor is a projection from ${\cal V}^{\otimes 4}$ into the spin-1/2 subspace. From the SU(2) conservation rules, one can only have an odd number, one or three, of virtual states carrying half-integer spins (which could be different like $1/2$ and $3/2$). Hence the SU($2$) tensor basis can be split into two sets, and from each of them one can build two separate families of U($1$)-symmetric tensors. However, combining basis tensors of the two sets, i.e. with different numbers of half-integer virtual spins, will break the U(1)-gauge symmetry into $\mathbb{Z}_2$. 

There is however some sort of fine-tuning (or ''protection'') in the set-up itself: this is in fact the form of the quench Hamiltonian which i) is SU(2)-symmetric and ii) involves only NN bonds. In the special case ii) one can use the simple TS decomposition in terms of 2-site gates. From i) each gate acting on a bond can only change the virtual spin $S$ into $S+1$ or $S-1$ and, hence, will not change the U(1)-gauge symmetry of the state. Such a property will no longer be true if either i) or ii) is broken. In particular, the gauge U($1$) symmetry will be spoiled by a small amount of disorder introducing (small) violations of i) or ii).

\begin{figure}[htb]
        \includegraphics[width=\columnwidth]{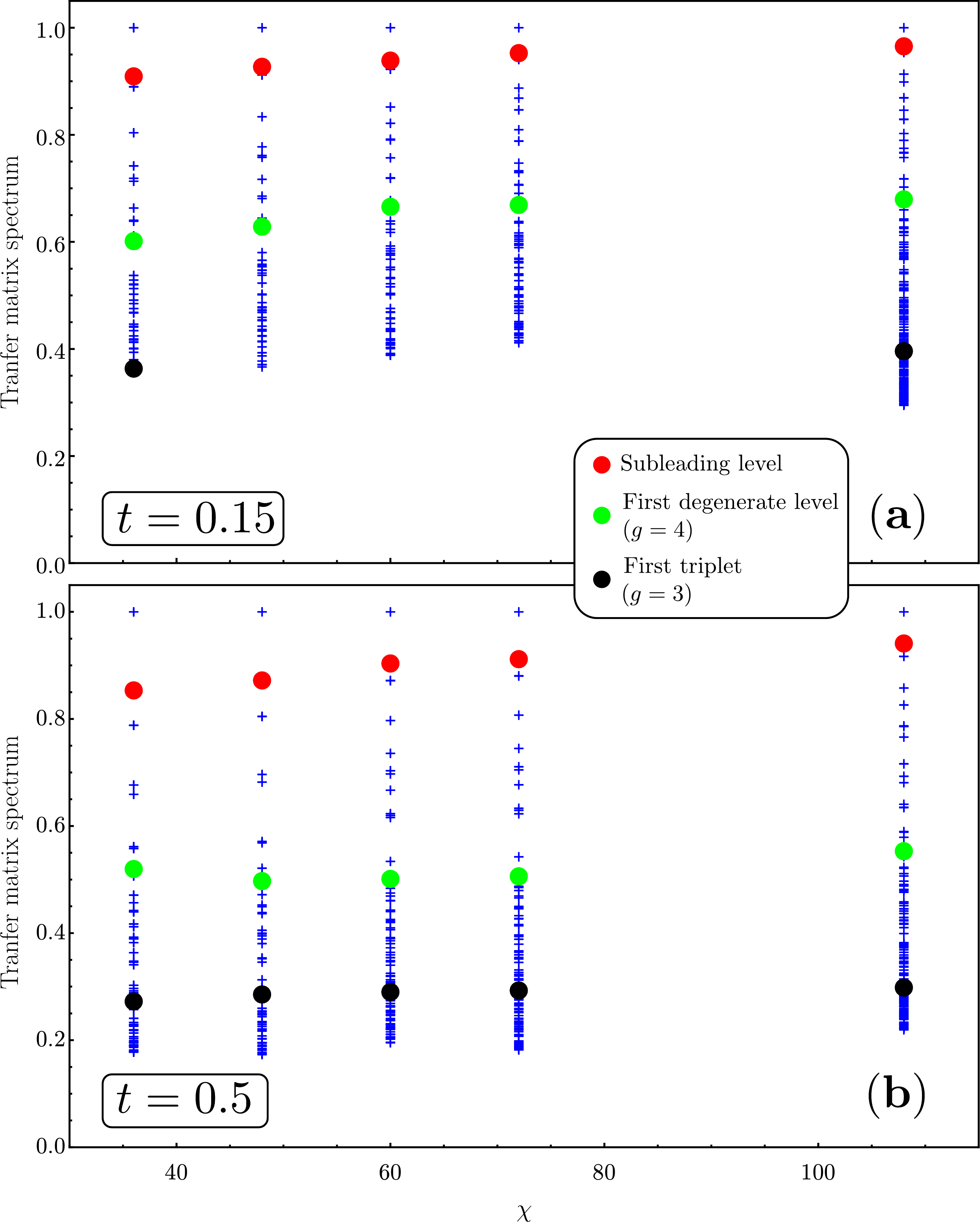}
    \caption{
    TM spectrum at a fixed value of $t=0.15$ (a) and $t=0.5$ (b), computed for several values of $\chi$. Different symbols are used to highlight particular levels, the subleading $g=1$ eigenvalue (red dot) and the largest eigenvalues $\lambda_{\rm spinon}$ and $\lambda_T$ with degeneracy $g=4$ (green dot) and $g=3$ (black dot), respectively. A gapless continuous singlet spectrum is expected in the $\chi\rightarrow\infty$ limit. 
    \label{fig:tm}}
\end{figure}

\begin{figure}[htb]
    \centering
    \includegraphics[width=\columnwidth]{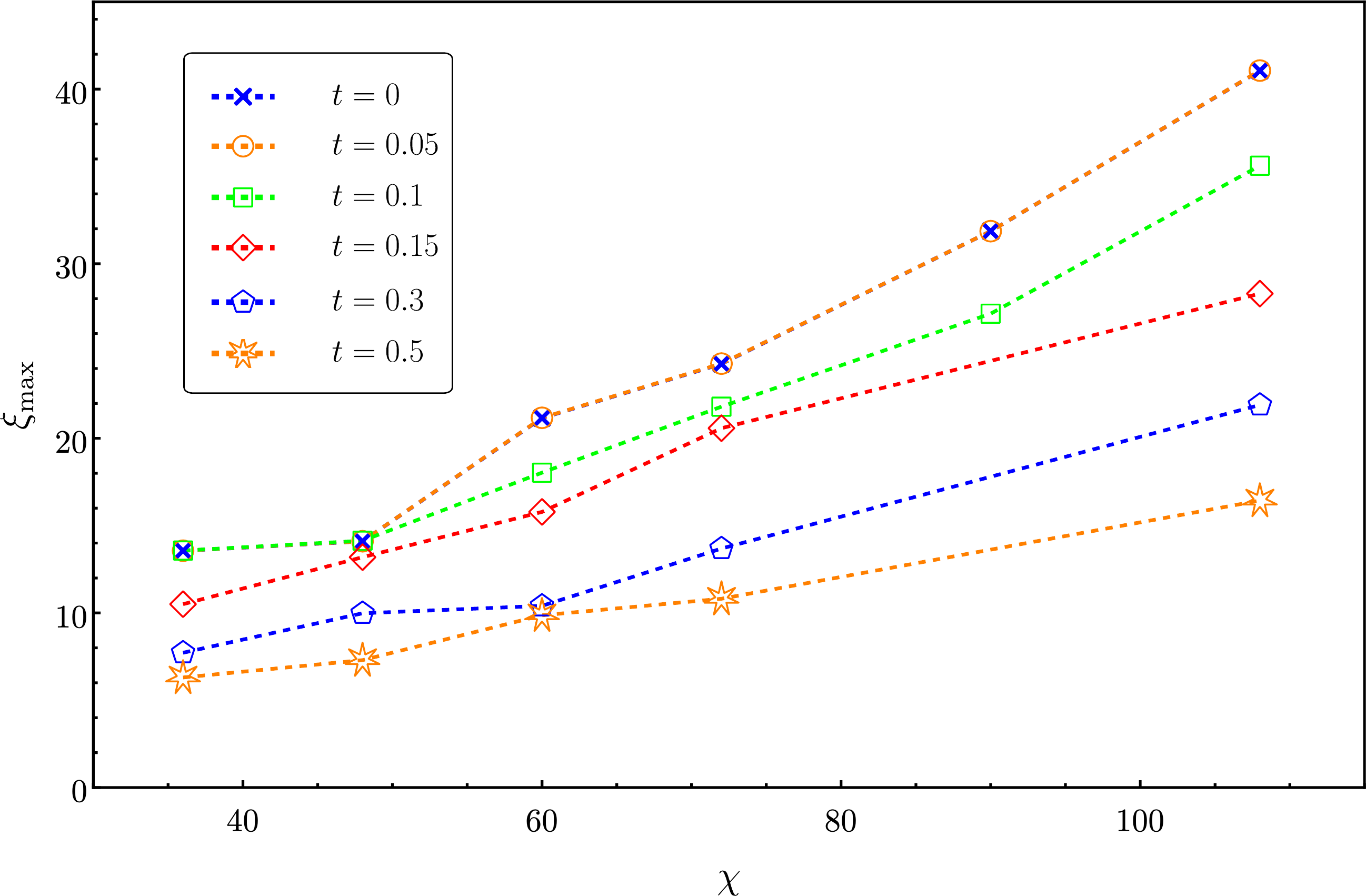}
    \caption{
    Maximum correlation length $\xi_{\rm max}$ for several values of time $t$ showing absence of saturation as a function of $\chi$.   
    \label{fig:maxcorr}}
\end{figure}

{\it Transfer matrix --}
To investigate further the expected critical nature of the state we have computed the spectrum of the transfer matrix (TM) build from two $T$ environment tensors~\cite{Chen2018a}. The (modulus of the) leading eigenvalues are shown in Fig.~\ref{fig:tm} for $t=0.15$ (a) and $t=0.5$ (b) (but similar results are also found for other values of $t$), normalizing the spectrum such that the largest eigenvalue is $\lambda_1=1$. The behavior of the spectra with increasing $\chi$ suggests a vanishing gap $\lambda_1-\lambda_2$ between the largest eigenvalue and the subleading one $\lambda_2$. The maximum correlation length $\xi_{\rm max}=-1/\ln{(\lambda_2/\lambda_1)}$ is then expected to diverge with $\chi$, as confirmed in Fig.~\ref{fig:maxcorr} showing a linear behavior of $\xi_{\rm max}$ with $\chi$ and no sign of saturation. In contrast to the gapless singlet ($g=1$) spectrum a gap is seen in the doublet (spin-1/2) and triplet ($g=3$, spin-1) sectors. Note that the doublet spectrum exhibits an extra two-fold degeneracy ($g=4$). Such results are consistent with short-range spin-spin correlations, as discussed later. 

\begin{figure}[htb]
    \centering
    \includegraphics[width=\columnwidth]{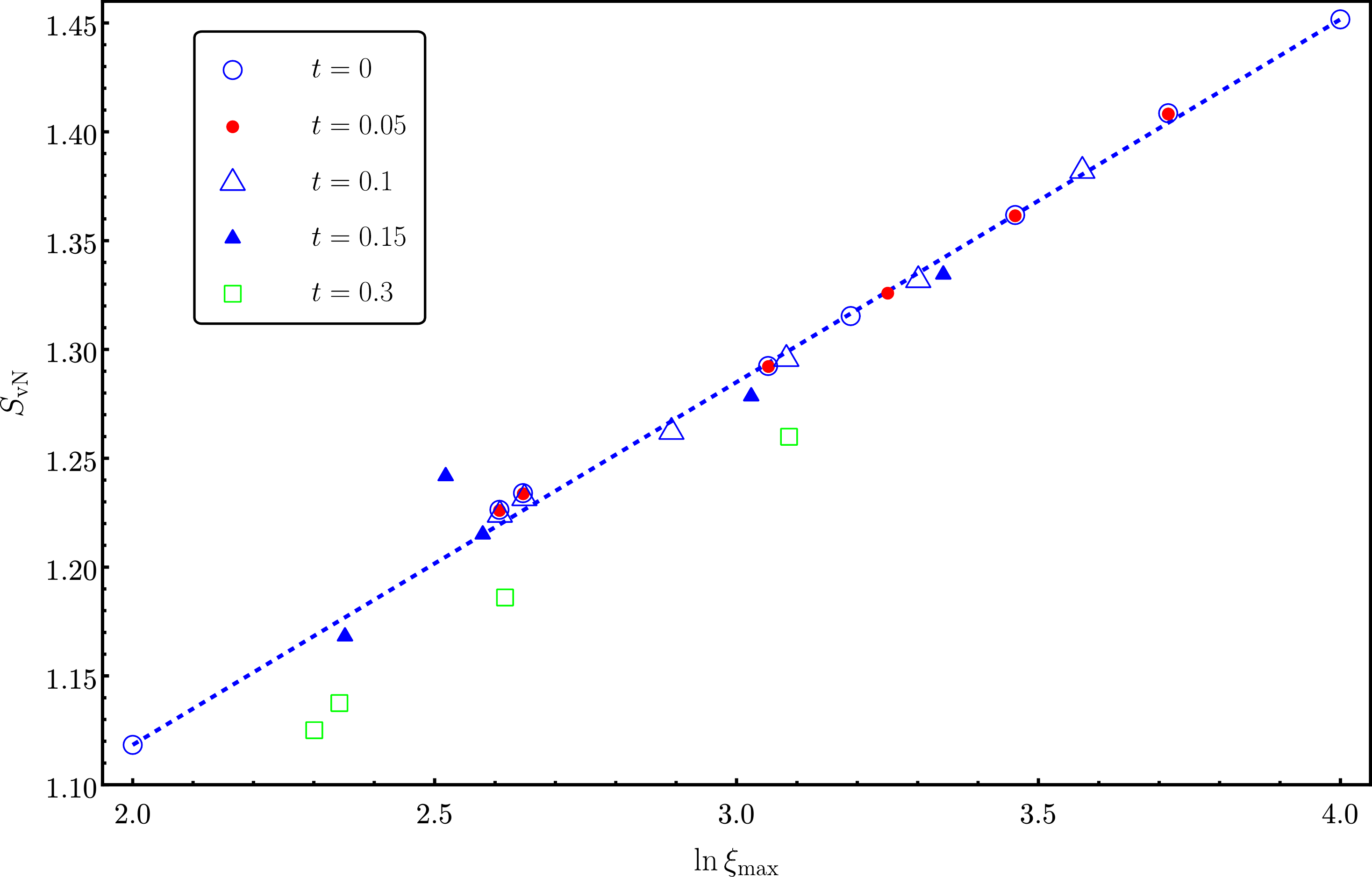}
    \caption{
     Entanglement entropy of the boundary MPS versus $\ln{(\xi_{\rm max})}$ for different time $t=0,0.05,0.1,0.15, 0.3$. The dotted line correspond to the behavior expected 
    for $c=1$. 
    \label{fig:Svn}}
\end{figure}

{\it Central charge --} From the CTMRG environment one can build the boundary MPS of bond dimension $\chi$ and ``physical" dimension $D^2$. The critical nature of the bulk PEPS is reflected in the critical nature of the boundary chain characterized by a central charge $c=1$ for all time $t$. We have confirmed this feature by computing the Von Neumann entanglement entropy $S_{\rm vN}$ of the MPS. The later is shown in Fig.~\ref{fig:Svn} as a function of $\ln{(\xi_{\rm max})}$ for small $t$ values. Fitting the data as $S_{\rm vN}\sim \frac{c}{6} \ln{(\xi_{\rm max})}$\cite{Calabrese_2004,Calabrese2009} one obtains $c\simeq 1$ consistent with the expected result. 

\subsection{Spin-spin correlations} 
\label{subsec:spin_corr}

\begin{figure}[htb]
    \centering
    \begin{subfigure}{}
    \includegraphics[width=\columnwidth]{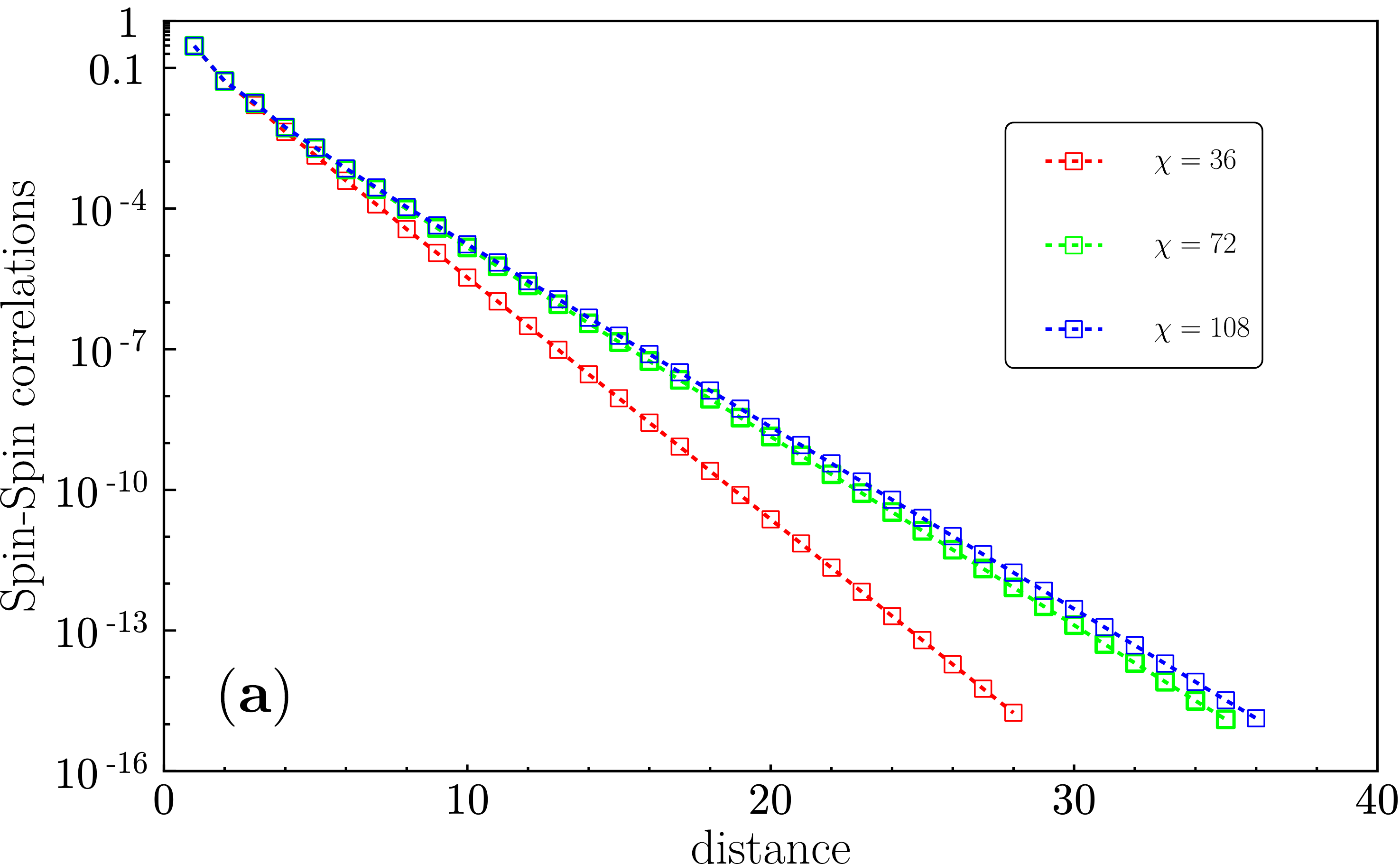}
    \end{subfigure}
    \begin{subfigure}{}
    \includegraphics[width=0.96\columnwidth]{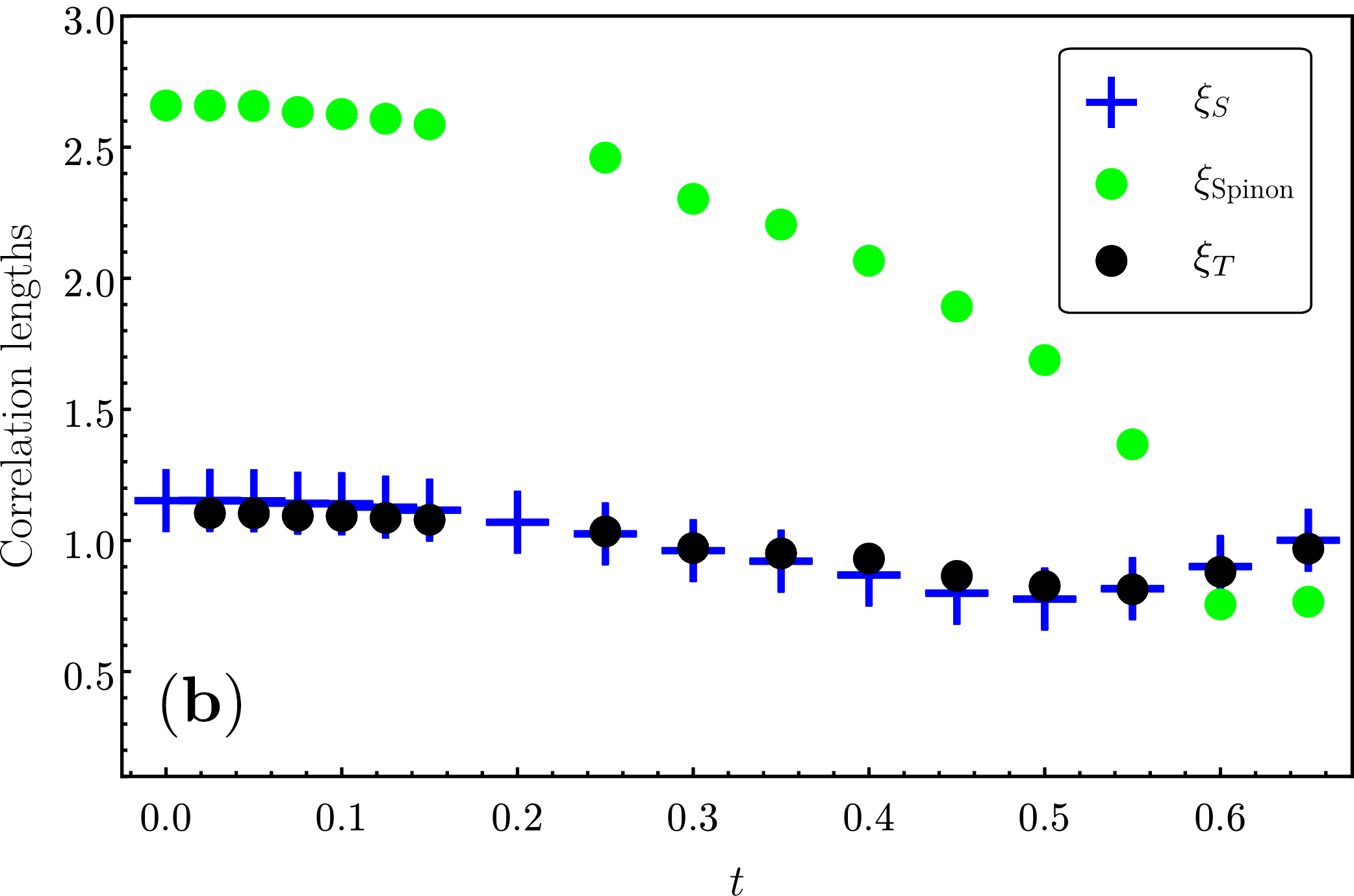}
    \end{subfigure}
    \caption{
    (a) Spin-spin correlation versus distance for $t=0.15$ in semi-log scale for several values of $\chi$. The linear fit corresponds to an exponential decay. (b) Correlation lengths at $\chi=108$ versus time $t$: $\xi_S$ extracted from fits of the spin-spin correlations; $\xi_T$ and $\xi_{\rm spinon}$
    extracted from the TM spectra shown in Fig.~\ref{fig:tm} (using same colors of dots). 
    \label{fig:spincorr}
    }
\end{figure}

The previous results suggest that the time-evolving state bears similar properties as the initial NN RVB state, although with rapidly growing entanglement. First, the diverging correlation length corresponds to  power-law decaying dimer-dimer correlations. Secondly, spin-spin correlations are expected to be short-range. This is indeed seen in Fig.~\ref{fig:spincorr}(a) for $t=0.15$. The corresponding spin-spin correlation length $\xi_S$ is shown in Fig.~\ref{fig:spincorr}(b) as a function of time $t$. Interestingly, the values extracted from fits of the spin-spin correlations match the values $\xi_T$ obtained from the TM spectra (see e.g. Fig.~\ref{fig:tm} (a,b) for $t=0.15$ and $t=0.5$): considering the leading spin-1 ($g=3$) eigenvalue $\lambda_T$, one gets $\xi_{T}=-1/\ln{(\lambda_T/\lambda_1)}\simeq \xi_S$. Fig.~\ref{fig:spincorr}(b) also shows the behavior of the spinon correlation length $\xi_{\rm spinon}=-1/\ln{(\lambda_{\rm spinon}/\lambda_1)}$ where $\lambda_{\rm spinon}$ is the leading eigenvalue with $g=4$ corresponding to two degenerate spin-doublets ($S=1/2$). Remarkably, we observe that all these correlation lengths vary moderately as a function of time. 

\section{Conclusions} %
\label{sec:conclusion}

In this work we have considered a simple quench setup to investigate the non-equilibrium time dynamics of a genuine critical spin liquid i.e. realizing a Coulomb phase. Our description uses the iPEPS algorithm to approximate the time evolution under the application of a NN Heisenberg interaction, allowing for the preservation of all symmetries, both the lattice symmetry (by considering a unique $C_{4v}$-symmetric site tensor) and the spin-rotation SU(2) symmetry directly encoded at the level of the site tensor. At every time step, the inerrant breaking of the point group $C_{4v}$ symmetry due to the successive application of four (non-commuting) two-site gates followed by SVD truncations is repaired by subtle gauge transformations on the four tensor virtual legs. 

Although our procedure is accurate only at small time due to the limited tractable bond dimension $D$, we argue that the observed stability of the critical nature of the evolving state is valid at all times in the case of {\it nearest neighbor} interactions. The robustness of the U(1) gauge symmetry, the intrinsic origin of the criticality, then suggests absence of ``thermalization". Extension of this work to an initial $\mathbb{Z}_2$ topological spin liquid and/or a longer-range (spin symmetric) interaction is left for future studies. 

\par\noindent\emph{\textbf{Acknowledgments---}} %
We acknowledge inspiring discussions with Ji-Yao Chen, Olivier Gauth\'e, Norbert Schuch, Luca Tagliacozzo, Laurens Vanderstraeten, Frank Verstraete and support from the TNTOP ANR-18-CE30-0026-01 grant awarded by the French Research Council. This work was granted access to the HPC resources of CALMIP center under the allocation 2017-P1231.

\clearpage

\appendix


\section{Implementation of $C_{4v}$ and SU(2) symmetries} 
\label{appendix:sym}

Here we provide details on the steps represented in Fig.~\ref{fig:su_details} (h-i). Once a uniform consistent gauge is fixed for the four virtual legs (see text), 
the evolved tensor is projected onto the symmetric basis $\{T_a\}$. In this process the resulting tensor becomes fully symmetric under $C_{4v}$ and SU(2), in particular the noise $\varepsilon$ introduced during the gauge fixing process is removed. 

\begin{figure}[htb]
	\centering
	\includegraphics[width=0.7\columnwidth]{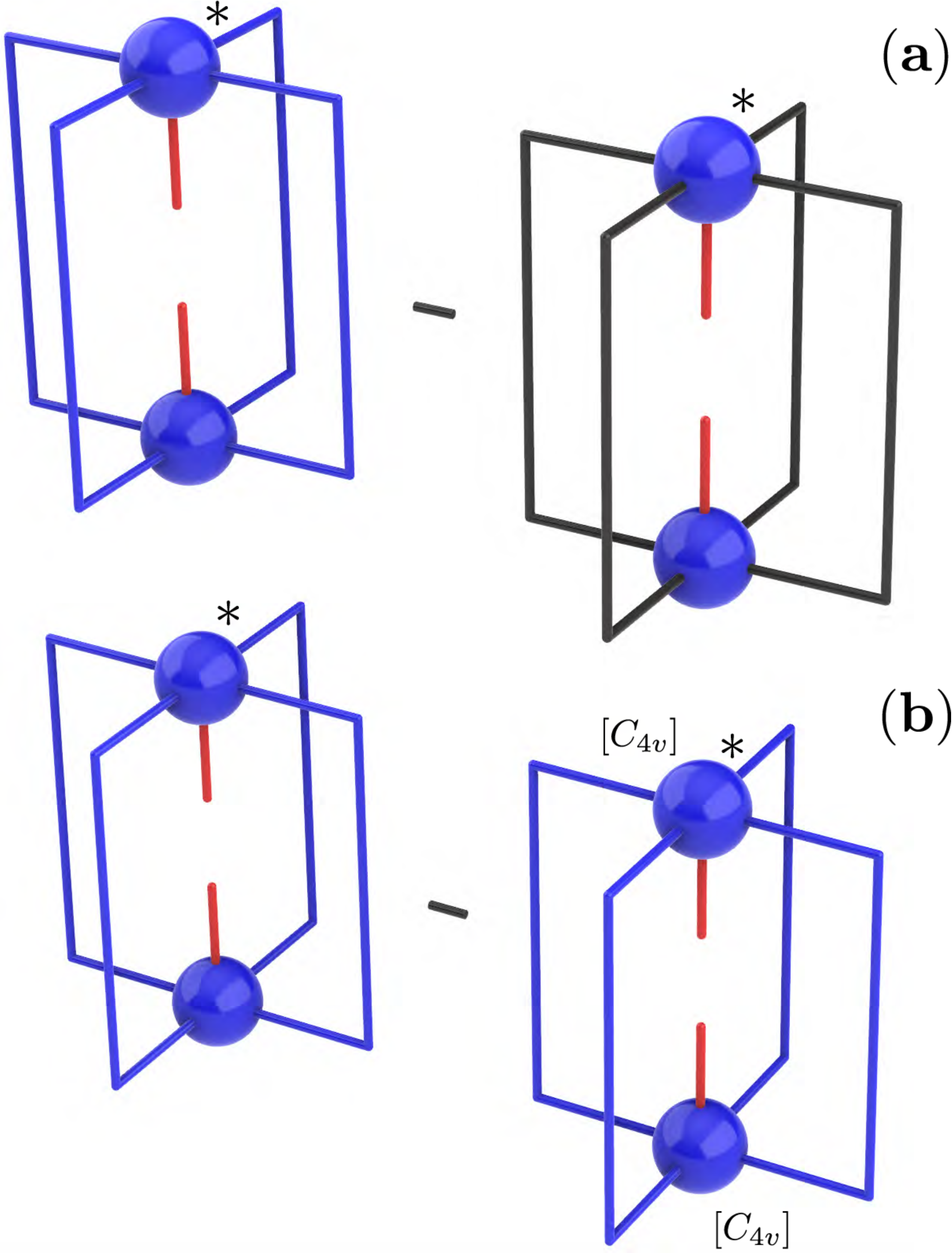}
	\caption{(a) Tensor increment $T_i$ and (b) Tensor dissymetry $T_d$ using the same definitions as in Fig.~\ref{fig:su_details}.
	}
	\label{fig:tits}
\end{figure}

In order to test the accuracy of this procedure, let us define two $d \times d$ matrices according to Fig.~\ref{fig:tits}:
\begin{itemize}
	\item The tensor increment $T_i$ evaluates the variation of site tensor under time evolution. From its definition as a bilayer, it is obviously gauge independent. Since ${\cal A}(t+\tau)-{\cal A}(t) = \alpha \tau +{\cal O}(\tau^2)$, the norm $|| T_i ||$ is expected to scale like $\tau$,
	\item Tensor dissymetry $T_d$ measures the effect of explicit $C_{4v}$ symmetrization. It is evaluated using the gauge fixed tensor, as $C_{4v}$ symmetrization would be meaningless otherwise. Corrections induced by symmetrization on ${\cal A}(t+\tau)$ are due to the non-commutativity of the 4 substeps and thus expected to occur at most at order $\tau^2$. Hence $||T_d||$ should scale as $\tau^2$.
\end{itemize}

\begin{figure}[htb]
	\centering
	\includegraphics[width=0.9\columnwidth]{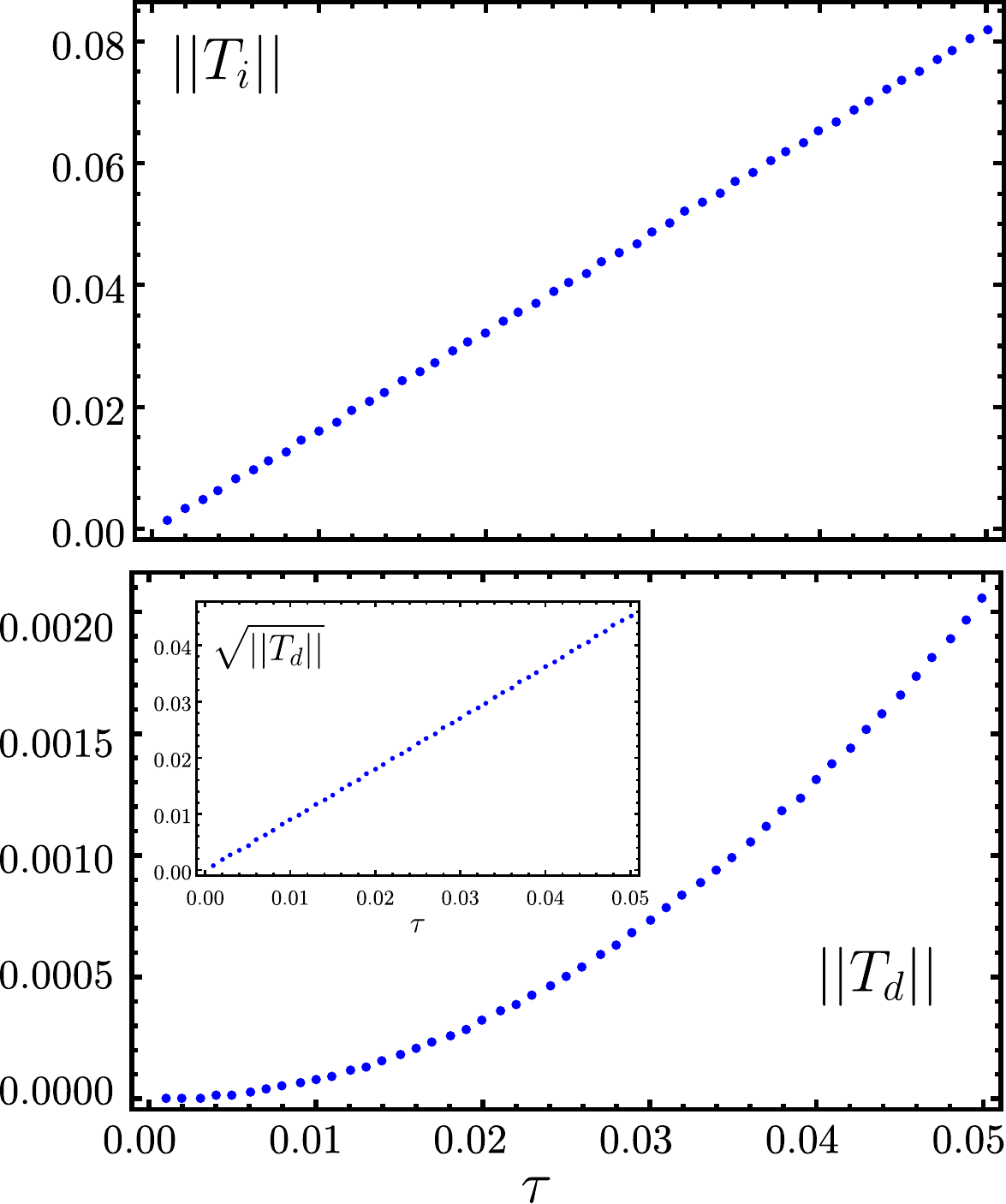}
	\caption{Small $\tau$ dependence of $||T_i||\sim 1.6287 \tau $ (upper panel) and $|| T_d||\sim 0.817375 \tau^2$ (lower panel). 
	}
	\label{fig:c4v}
\end{figure}

We checked these scalings using ${\cal V}=0\oplus\frac{1}{2}\oplus 1$ (the justification for using this ansatz was given in section~\ref{subsec:virtual}). Figure~\ref{fig:c4v} shows that indeed $|| T_i || \sim \tau $ and $|| T_d || \sim \tau^2 $. Hence it is always possible to choose $\tau$ small enough such that $|| T_d || \ll || T_i ||$. It is typically the case for $\tau=0.025$ used in the following.

In a final stage (see Fig~\ref{fig:su_details} (i)), the tensor is projected in the SU(2) symmetric basis $\{T_a\}$. This basis being orthogonal and normalized, the error of this projection is simply evaluated as $1-\sum_a |\langle {\cal A }(t)| T_a \rangle|^2$ and we checked that it never exceeds $10^{-10}$ for $\tau=0.025$.

\section{Factorisations of symmetric PEPS
\label{appendix:factorisation}}

In this section, we briefly describe the matrix factorizations that are frequently used in the current work. In particular, we are interested in factorizing a complex symmetric matrix into a diagonal form i.e., 
\begin{equation}
    A = V D V^T
\end{equation}
where $D$ is a diagonal matrix and $V$ is an invertible matrix. We also explain how these matrices can be used to truncate the matrix $A$.

\subsection{Autonne-Takagi Decomposition}
\label{appendix:takagi}
The Autonne-Takagi factorization of a complex symmetric  matrix $A$ is defined as,
\begin{equation}
    A  = US U^T
\end{equation}
Where $S$ is a diagonal matrix with positive real entries called the  singular values of the matrix A and U is a complex unitary. In order to find the unitary $U$, we follow the procedure described in Ref. \cite{CHEBOTAREV2014380}.

For any complex matrix, we define the singular value decomposition as,
\begin{equation}
\label{simpleSVD}
    A = U S V^\dagger    
\end{equation}
where $S$ is the diagonal with singular values and $U$ and $V$ are unitary matrices with corresponding singular vectors and $\dagger$ denotes Hermitian conjugate.

We define the unitary matrix $Z = U^\dagger V^*$ where $V^*$ denotes the complex conjugate of $V$. Since $A$ is complex symmetric, we can rewrite the SVD as,
\begin{eqnarray}
&U S V^\dagger = &A = A^T = V^* S U^T \nonumber\\
&\Rightarrow& SV^\dagger U^* = U^\dagger V^*  S \nonumber\\
&\Rightarrow& ZS=SZ^T \nonumber \\
&\Rightarrow&  ZS Z^* = S \nonumber
\end{eqnarray}

Since the above equation preserves the spectrum of S, it is a similarity transformation. Hence, $Z$ and $Z^*$ should be multiplicative inverse of each other. i.e. $Z^\dagger= Z^*\Rightarrow\quad Z = Z^T$. This implies that $Z$ and $S$ commute with each other. 
\begin{eqnarray}
&(ZS)_{ij} = &z_{ij}S_{jj} \; \text{and} \; (SZ)_{ij} = S_{ii}z_{ij}\nonumber\\
&\Rightarrow& (z_{ij}S_{jj} - S_{ii}z_{ij})  = 0 \nonumber \\
&\Rightarrow& z_{ij}(s_{i} - s_{j}) = 0 \nonumber \\
&\Rightarrow& z_{ij} =0\quad 	\forall s_{i} \neq s_{j} \nonumber
\end{eqnarray}

Hence, we note that any such matrix $Z$ that commutes with a diagonal matrix $S$ should be diagonal if all the entries in $S$ are distinct. If $S$ has repeated singular values, $Z$ has a block diagonal structure with the size of the blocks equal to the degeneracy of the singular values i.e., $Z = \bigoplus_{k} B_{n_k \times n_k}$ where $n_k$ is the multiplicity of the $k$-th distinct singular value and $B_{n_k \times n_k}$ is the corresponding block diagonal part of $Z$. Similarly, we can rewrite $S = \bigoplus_{k} s_k\cdot {\mathbbm{1}}_{n_k \times n_k}$. Since every block $B_{n_k \times n_k}$ commutes with the identity matrix, the overall matrix too commutes with the diagonal matrix $S$. Since any power of $Z$ can also be written in similar block diagonal form, they too commute with the diagonal matrix $S$. Hence we can write, $Z^{\frac{1}{2}}S = SZ^{\frac{1}{2}}$.
\begin{eqnarray}
\label{Zn2}
    A &=& U S V^\dagger \nonumber \\
    &=& U S V^\dagger U^* U^T \nonumber \\
    &=& U S Z^T U^T \nonumber \\
    &=& U (Z^{\frac{1}{2}})^T S (Z^{\frac{1}{2}})^T U^T \nonumber \\
    &=& U_Z S U_Z^T
\end{eqnarray}

We therefore get $A = U_Z S U_Z^T$ with $U_Z = U (Z^{\frac{1}{2}})^T$. Note that the resultant unitary $U_Z$ is not unique, since the singular matrices $U$ and $V$ themselves are not unique. Let  two such unitaries $U_{Z_1}$ and $U_{Z_2}$ be related by $U_{Z_1} = U_{Z_2} \phi$, where $\phi$ is a unitary matrix, then,
\begin{eqnarray}
\label{eq:takagi_gauge}
    A &=& U_{Z_1} S (U_{Z_1})^T  \nonumber \\
    &=& U_{Z_2} (\phi S \phi^T) (U_{Z_2})^T \nonumber \\
    &=&  U_{Z_2}  S  (U_{Z_2})^T
\end{eqnarray}
i.e. $(\phi S \phi^T) = S$. Since this transformation preserves the spectrum of S, it has to be a similarity transformation, i.e. $\phi \phi^T = I$. This also implies that the matrix $\phi$ has to commute with $S$. Hence, if the singular values $S$ are all distinct, the matrix $\phi$ is a orthogonal diagonal matrix i.e. it's a diagonal matrix with only $\pm1$ as its diagonal entries. If $S$ has multiplicities, the matrix $\phi$ can take a block diagonal structure with $n_k \times n_k$ orthogonal blocks where $n_k$ is the multiplicity of the corresponding singular value $S_k$.

\subsection{Orthogonal Decomposition}
If the complex symmetric matrix $A$ is diagonalizable, one can diagonalize it by using a set of eigenvectors i.e.,

\begin{eqnarray}
    A E &=& E D  \nonumber \\
    \Rightarrow A &=& E D E^{-1}
\end{eqnarray}
where $D$ is a diagonal matrix whose entries are the eigenvalues of $A$ and E is the matrix with the corresponding eigenvectors~\cite{strang09}.
 Now, we use the symmetry argument $A=A^T$,
\begin{eqnarray}
    &A = &EDE^{-1} = A^T = (E^{-1})^T D E^T \nonumber\\
    &\Rightarrow& ED = (E^{-1})^T D E^TE  \nonumber\\
    &\Rightarrow& ZD = DZ, \nonumber
\end{eqnarray}

where we have defined $Z=E^TE$.
Note that eigenvectors of different eigenvalues of any complex symmetric matrix are always orthogonal (not orthonormal). Indeed, let $u$ and $v$ be two eigenvectors corresponding to two different eigenvalues $\lambda$ and $\mu$, then,
\begin{equation}
    u^T \cdot Av - (u)^T A^T \cdot v =  (\mu - \lambda)(u^T \cdot v) 
\end{equation}
If the matrix $A$ is symmetric, the LHS of the equation should be zero, and since $\lambda$ and $\mu$ are distinct, the dot product $u^T \cdot v$ should be zero which can only happen when $u$ and $v$ are orthogonal.
Hence the matrix $Z=E^TE$ is a diagonal matrix if the A has nonrepeating distinct eigenvalues. However, in the presence of multiplicities, the matrix exhibits a block diagonal form. Unlike Takagi decomposition, this block diagonal form has to do with the absolute values of the diagonal entries D. Eigenvalues which have the same absolute value but are a complex phase away(like the complex conjugate) can still have the same eigenvectors because if $u$ is an eigenvector, the vector $e^{i\phi} u$ is also an eigenvector.
Following the same arguments as in the previous section, we get $Z^{\frac{1}{2}}D = DZ^{\frac{1}{2}}$. Notice that the square root of Z should also have the same block diagonal form as Z. Then, we get
\begin{eqnarray}
    A &=& EDE^{-1} \nonumber \\
    &=& EDE^{-1}(E^T)^{-1}E^T \nonumber \\
    &=& EDZ^{-1}E^T \nonumber \\
    &=& EZ^{-1/2}DZ^{-1/2}E^T \\
  &=& (EZ^{-1/2})D(EZ^{-1/2})^T \nonumber \\
  &=& ODO^T \, \nonumber.
  \label{SimpleOrthogonal}
\end{eqnarray}
Thus the matrix $O = EZ^{-1/2} = E(E^TE)^{-1/2}$ is a set of orthogonal (complex) eigenvectors of A. Note that in practice, eigenvectors of eigenvalues which are distinct but whose absolute values are quite close can generate very similar eigenvectors. 

The major difference between the Autonne-Takagi decomposition and the orthogonal decomposition is that, since the orthogonal decomposition results in orthogonal projectors, the operation $O^T B O$ is a similarity transformation on the matrix B whereas $U^T B U$ is not. Both methods can be used to  a factorize any symmetric matrix into a product of a matrix and its transpose, i.e. $ A = M M^T$ where $M = OD^{1/2}$ or $M = US^{1/2}$. 

\subsection{Relation between Singular values and Eigenvalues}

Since both methods would give us a decomposition of the matrix into a product of a matrix and it's transpose, we can try to establish a relation between them by comparing the resultant matrices. Since the number of non-zero elements of $S$ and $D$ is a equal to the rank of the matrix, the size of non-zero elements of $S$ and $D$ should be the same.
Let us write $A = USU^T = U\sqrt{S}\sqrt{S}U^T = (U \sqrt{S}) (U \sqrt{S})^T$ or, 
similarly, $A = ODO^T = O\sqrt{D}\sqrt{D}O^T = (O \sqrt{D}) (O \sqrt{D})^T$.
The identification of the two forms, 
\begin{equation*}
   (U \sqrt{S}) (U \sqrt{S})^T  = (O \sqrt{D}) (O \sqrt{D})^T \, ,
\end{equation*}
implies that
\begin{equation*}
  (U \sqrt{S}) \phi  = (O \sqrt{D}) 
\end{equation*} where $\phi$ is a complex orthogonal matrix. Hence,
\begin{equation*}
  O = U\sqrt{S}\phi \sqrt{D}^{-1}
\end{equation*} 
Making use of $O^TO = I$, we then obtain
\begin{eqnarray}
 \label{equal}
   &(U\sqrt{S}\phi \sqrt{D}^{-1} )^T (U\sqrt{S}\phi \sqrt{D}^{-1} ) = I \nonumber \\
   \Rightarrow &\sqrt{S}U^TU \sqrt{S} = \phi D \phi^T 
\end{eqnarray} 
Comparing the diagonal elements in LHS and RHS of Eq. \ref{equal}, we get,
\begin{eqnarray}
\label{equal2}
    C_{ii} &= \sum_k\sqrt{s}_{i} U_{ik}^TU_{ki}\sqrt{s_{i}} = \sum_k d_{k}\phi_{ik}\cdot\phi_{ki}^T \nonumber \\
    &\Rightarrow \sum_{i=1}^k d_{i} = \sum_{i=1}^k s_{i} (U_{i}^T \cdot U_{i})
\end{eqnarray}
Hence, it is clear that the eigenvalues and singular values are equal if and only if the unitary $U$ is purely real. Additionally, if the singular values are degenerate, the pre-factor term $U^T_{i}U_{i}$ need not be the same. Hence, $D$ might not always have the exactly same multiplet structure of $S$. This is particularly important while truncating the matrix. Since we do not want to cut through the multiplets, we should be more lenient with cutting through multiplets of $D$, often grouping values that are slightly off as degenerate. Additionally, we have,

\begin{equation*}
   \sum_i|d_{i}| =  \sum_i |s_{i}| |(U_{i}^T U_{i})| \leq \sum_is_i ||U_i|| \leq  \sum_i s_i
    \label{equal3}
\end{equation*} 
Thus, since sum total of magnitude of the first $n$ eigenvalues is lower than the sum of first $n$ singular values, we can infer that the eigenvalue spectrum decays quickly when compared to the singular values.

\subsection{Truncation}
Matrix truncation or dimensionality reduction is a transformation in which a matrix is projected from a higher dimensional space to a lower dimensional space such that the resultant matrix retains the maximum information and properties. In other words, for a given square matrix $M_{p\times p}$ we find a matrix that is lower order $N_{q\times q}$ for a given $q<p$ such that the matrix $N_{q\times q}$ can replace $M_{p\times p}$ with as little change as possible.
\newline
In order to truncate a matrix, we have to find an appropriate projector $U_{p \times q}$ which projects the linear vector space of dimension $p$ to the vector space of dimension $q$, i.e.,
\begin{equation}
    U^{T}  M_{p\times p}  V^{} = N_{q\times q}
\end{equation}
Since we are dealing with complex symmetric matrices, we can assume without the loss of generality that if the resultant matrix is a complex symmetric, the projectors U and V are identical.
As we project the matrix to a lower dimensional space, there is a loss of information due to the truncation. We quantify this by bringing the matrix $N_{q\times q}$ back to the vectors space of $M_{p\times p}$ and take the norm of the difference of the resultant matrix from the original. This reconstructed matrix $\Tilde{U} N_{q \times q} \Tilde{U}$ is called a {\it Low Rank Approximation} of the original matrix $M$~\cite{Eckart1936}.

\begin{equation}
    {\text{Loss}}  = ||M - \Tilde{U} N \Tilde{U^T}||
\end{equation}
where $\Tilde{U}$ is the pseudo-inverse of matrix $U$. Note that from here on, the word norm shall be used to refer the Frobenius norm of a matrix. Ideally, in a tensor contraction, we apply $\Tilde{U}U$ on every bond and by absorbing them to the corresponding tensors, we reduce the dimension of the bond.

\subsection{Loss in Autonne-Takagi truncation}
In Autone-Takagi factorization, we use the complex conjugate of the unitary $U^{\dagger}$ obtained by the factorization to diagonalize the matrix i.e. $M = U S U^T$. We then truncate the resultant matrix by keeping the $q$ of the diagonal matrices and setting the rest to zero. Let us denote the truncated diagonal matrix by $\Tilde{S}(q)$.
\newline

In this case, the loss due to truncation is given by,
\begin{eqnarray}
{\text{Loss}} &=&  ||M - U \Tilde{S}(q) U^T|| = ||U (S - \Tilde{S}(q) )U^T|| \nonumber\\
    &=& ||(S - \Tilde{S}(q) )|| = \sqrt{\sum_{k=q+1}^p s_{k}^2}
\end{eqnarray}
    
Hence, to minimize this loss, we retain the singular values that are largest in magnitude. Once the appropriate values of $S$ which have to be deleted are determined, we remove the corresponding columns in the matrix $U$ to get the required projector.

\subsection{Loss in Orthogonal truncation}
In case of orthogonal decomposition, we use the orthogonal matrices obtained by Eigenvalue decomposition in Eq.~\ref{SimpleOrthogonal}. Since the resultant diagonal entries can be complex, we truncate by retaining the $q$ values that have the largest magnitude.

Estimation of loss due to truncation is tricky in case of complex matrices that are not Hermitian. Even in complex symmetric case, we can never get a exact rule estimating the loss. Hence, we try to find an upper bound for the error. By proving that the upper bound reduces with increase in the final dimension $q$, we comment that the quality of truncation, in general, increases with the dimension.

\begin{figure}
    \centering
    \includegraphics[width=\columnwidth]{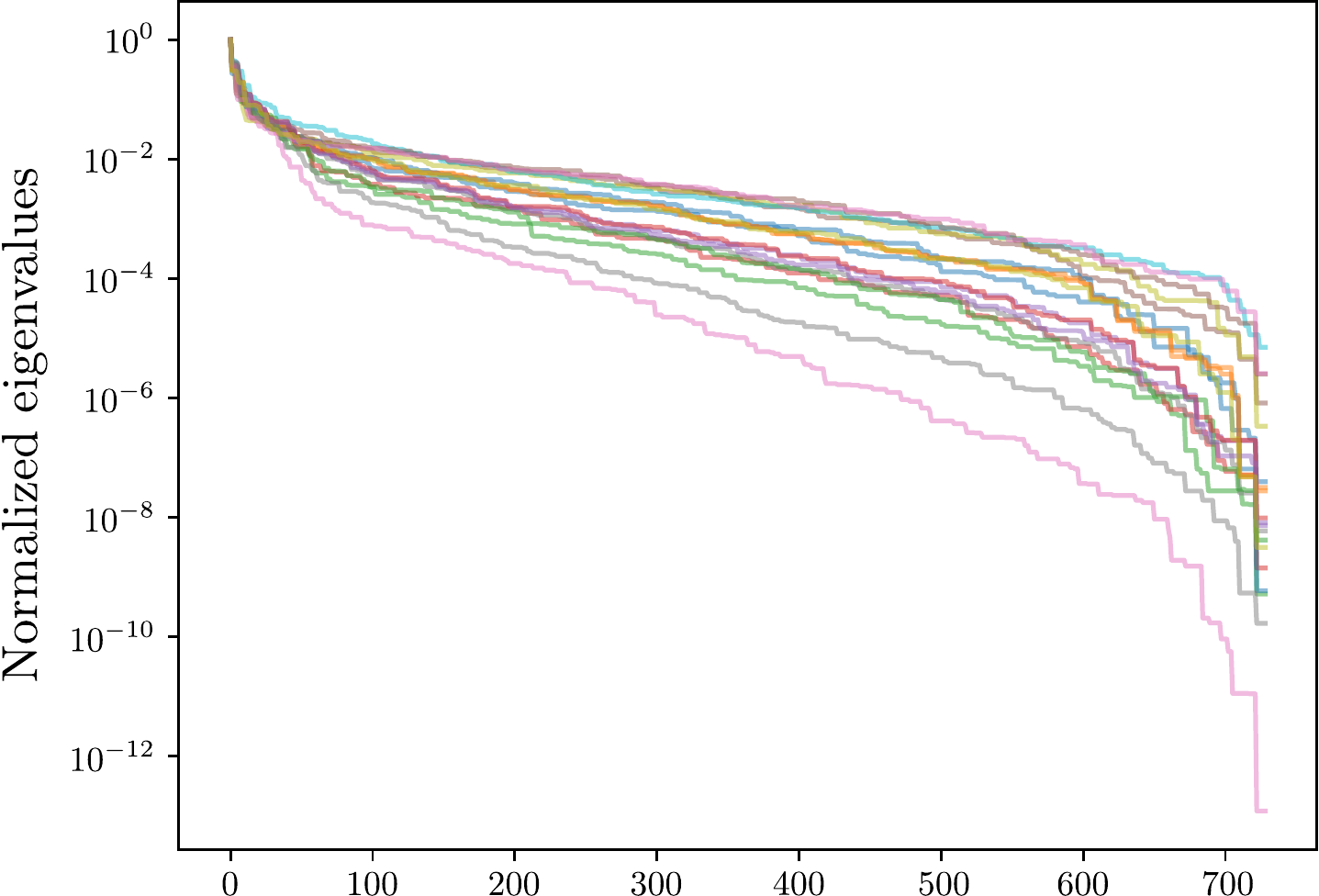}
    \caption{Normalized eigenvalues (in log scale) of $720\times 720$ corner matrices obtained from random PEPS ans\"atze~given by Eq.~\ref{eq:expansion} with bond dimension $D=6$.}
    \label{truncation_ortho1}
\end{figure}
\begin{figure}
    \centering
    \includegraphics[width=\columnwidth]{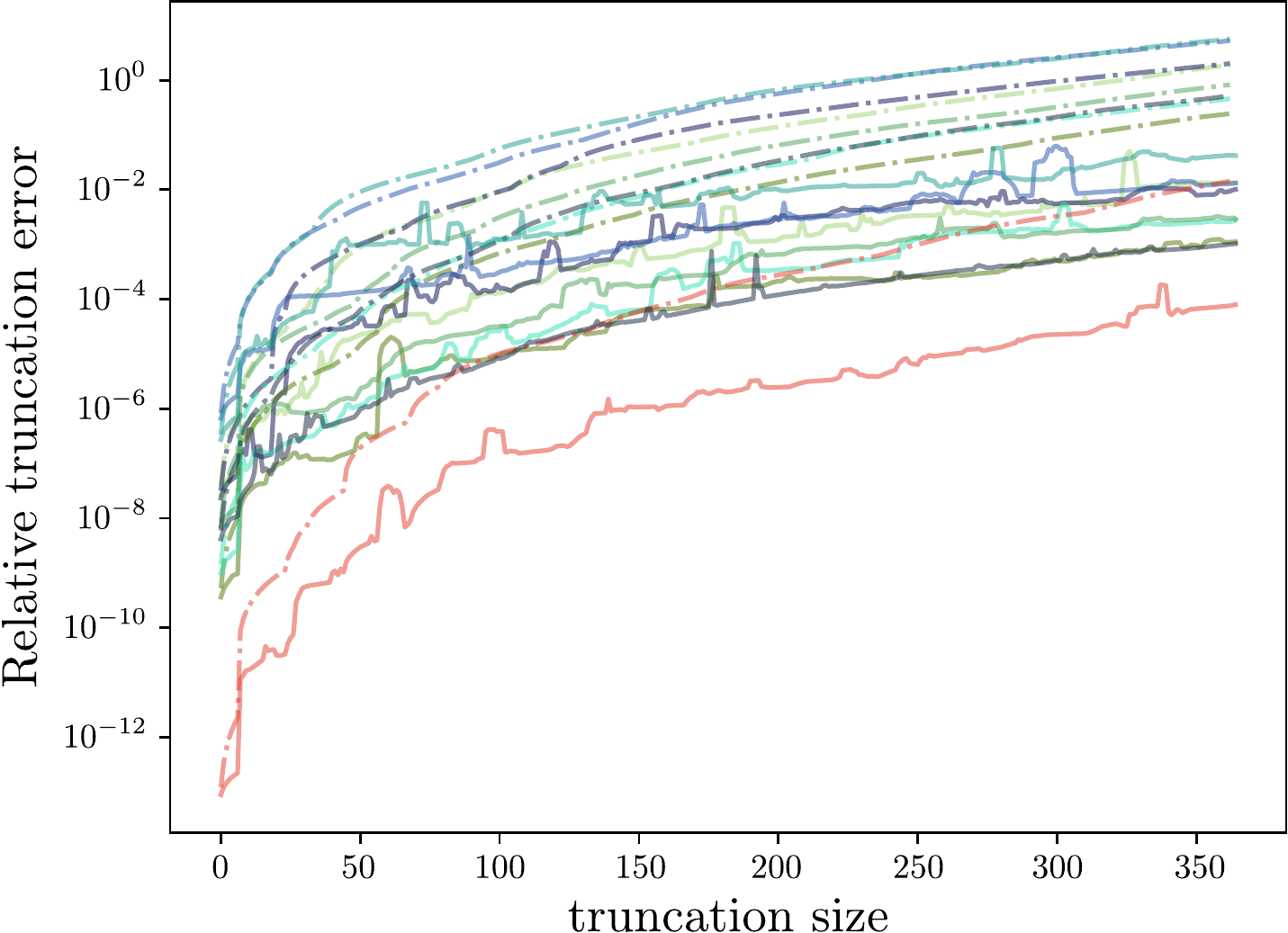}
    \caption{Relative truncation error (in log scale) of $720\times 720$ corner matrices obtained from random PEPS ans\"atze~given by Eq.~\ref{eq:expansion} with bond dimension $D=6$. The dotted lines indicate the upper bound calculated by Eq. \ref{orthoupperbound} while the corresponding solid lines indicate the actual errors. Notice that the general trend of error decreasing with decrease in truncation size.}
    \label{truncation_ortho2}
\end{figure}
In order to get an upper bound, we make use of the inequality that the Frobenius norm of the product of two matrices is lesser than or equal to the product of the norms of the individual matrices i.e., $||AB|| \leq ||A||||B||$ and the fact that a matrix and its transpose have the same entries and by extension, the same Frobenius norm.
From Eq. \ref{SimpleOrthogonal}, we Let us denote the truncated diagonal matrix by $\Tilde{D}(q)$ and the matrix with discarded values by $\Delta(D)$. Thus the loss due to truncation is given by,
\begin{eqnarray}
{\text{Loss}} &=&  ||M - O \Tilde{D}(q) O^T|| = ||O \Delta(D) O^T||\nonumber\\
&=& ||O \sqrt{\Delta(D)} (O \sqrt{\Delta(D)})^T|| = ||PP^T||\leq ||P||^2\nonumber\\
\Rightarrow {\text{Loss}} &\leq& || O \sqrt{\Delta(D)}||^2 \leq ||O||^2||{\Delta(D)}||.
\end{eqnarray}

Now, using the fact that $\Delta(D)$ has only the $(p-q)$ smallest eigenvalues, and the rest set to zero, the $O$ in the above equation can be replaced by only the corresponding $(p-q)$ orthogonal eigenvectors i.e.,  
\begin{equation*}
    O\Delta\sqrt{D(q)} = O_{p\times (p-q)} \sqrt{\Delta{D(q)}_{(p-q)\times (p-q)}}
\end{equation*} Let us call this truncated $O$ with $(p-q)$ eigenvectors as $O_{reduced}$. Now since, $O$ comprises of orthogonal vectors,
\begin{eqnarray}
     p-q &=& ||I_{(p-q)\times (p-q)}|| \nonumber \\
     &=&||O_{\text{reduced}}O^T_{\text{reduced}}|| \leq ||O_{\text{reduced}}||^2 \nonumber\\
\Rightarrow ||O_{\text{reduced}}|| &\geq& \sqrt{(p-q)}
\end{eqnarray}
Hence, we get,
\begin{equation}
    {\text{Loss}} \leq {(p-q)}||\Delta(D)|| = (p-q) \sqrt{\sum_{k=q+1}^p |d_k|^2}
    \label{orthoupperbound}
\end{equation}

For a given matrix of size $p$, the upper limit of the loss due to truncation will go down with increase in the size of the truncated matrix $q$. 

Notice that this upper limit doesn't neccesarily imply that the error will always keep decreasing with increase in $q$. as shown in Fig. \ref{truncation_ortho1} and Fig. \ref{truncation_ortho2} where we plot the eigenvalues of 25 random $729 \times 729$ matrices obtained from initial corners of CTMRG process. We can observe that the general trend of truncation error goes down with the decrease in the magnitude of truncation, though it is not always strictly decreasing.

\subsection{Preferences}
We shall now comment the usability of the above mentioned decompositions in a given problem. It is clear from the truncation error that for any given case, the singular value decomposition will have a much accurate low-rank approximation. Hence, in any dimension reduction problems that involve a single isolated matrix (or tensor), one has to always use singular value decomposition. Examples include dimension reduction in the simple-update procedure~\ref{sec:num}.

Orthogonal decomposition is preferred in cases where one has to reduce the bond of a tensor that is a part of closed loops or in cases where the overall trace has to be minimized. In cases where the tensor contraction can be reduced to the trace of product of two matrices, the trunction has to be done by inserting an isometry on the legs of the bonds that are to be truncated. Since we are dealing with complex symmetric matrices, for the resultant isometries to preserve the complex symmetric nature of the matrices, the projectors forming the isometry have to be complex symmetric, for which one has to resort to the orthogonal decomposition. In this work, we use orthogonal decomposition to renormalize the corner matric and the edge tensor in the CTMRG procedure.

\begin{figure}
    \centering
    \includegraphics[width=0.9\columnwidth]{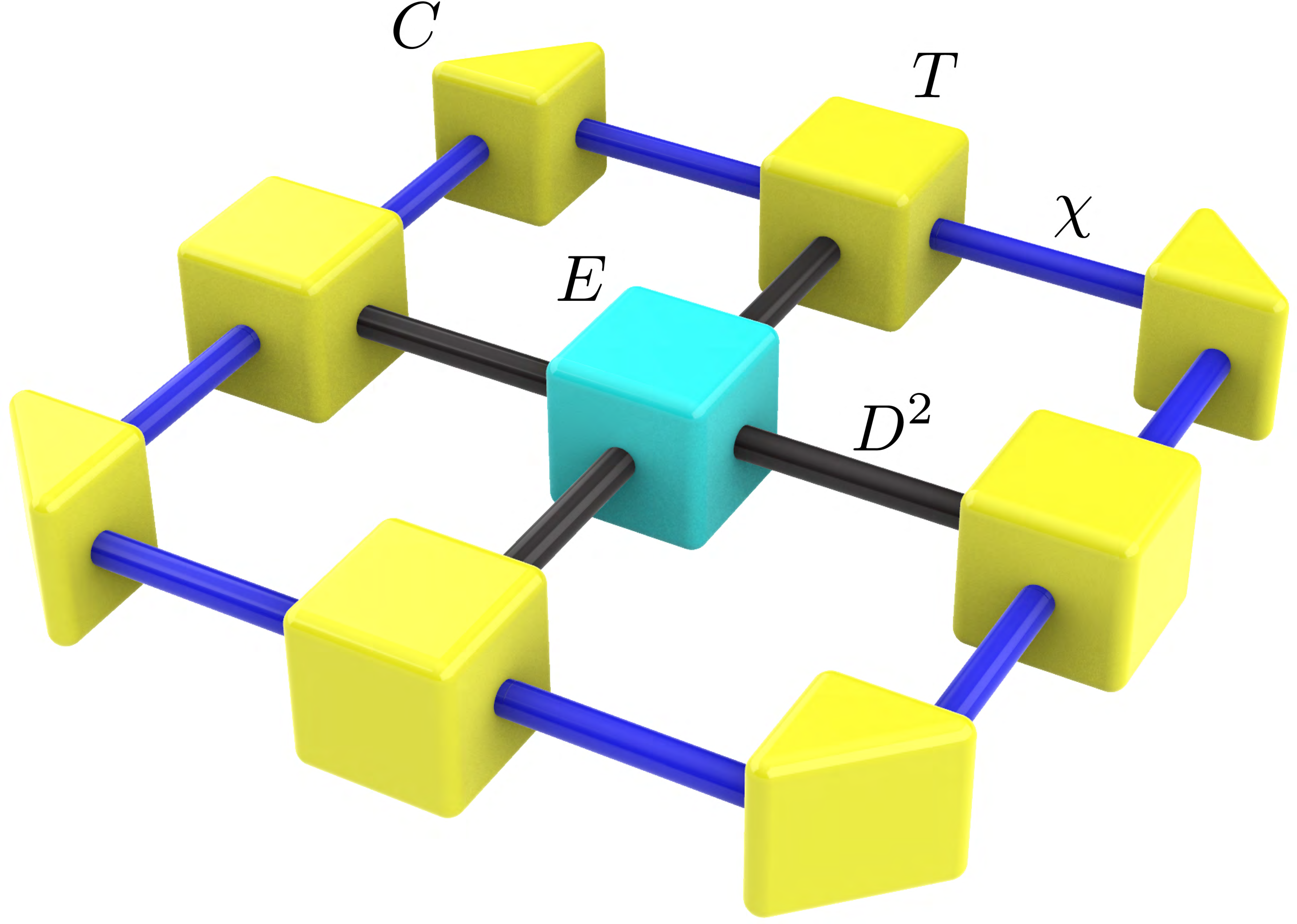}
    \caption{Single site symmetric environment obtained from the CTMRG procedure.}
    \label{fig:CTM1}
\end{figure}
\section{CTMRG
\label{appendix:CTMRG}}
In this section, we provide a description of the CTMRG procedure used to contract the iPEPS states used in this work (in order to compute observables like energy, correlation functions, etc...). The details of the CTMRG are similar to the one described in reference \cite{Orus2012}. More specifically its translationally invariant single-site $C_{4v}$-symmetric version~\cite{Mambrini2016}. In this scheme, when we contract all the active bonds of the bi-layer tensors in the thermodynamic limit, it leads to a $C_{4v}$-symmetric SU(2) environment with adjustable bond dimension $\chi$ as shown in Fig.~\ref{fig:CTM1}. The environment consists of a corner matrix $C$ of dimension $\chi \times \chi$ and a rank-3 transfer tensor $T$ of dimension $\chi \times \chi \times D^2$. However, the major difference is that the corner in this work is no longer Hermitian but complex symmetric. The edge tensor $T$ too is complex and exhibits reflection symmetry along the axis through its $D^2$ leg.
To reach the fixed point tensors, first $C$ and $T$ are initialized by contracting the corresponding legs of ket and bra tensor ${\cal A}^\dagger{\cal A}$. We then proceed to apply the renormalization procedure which is continued till the environment converges to a fixed point.\\
\begin{figure*}
    \centering
   \includegraphics[width=0.9\textwidth]{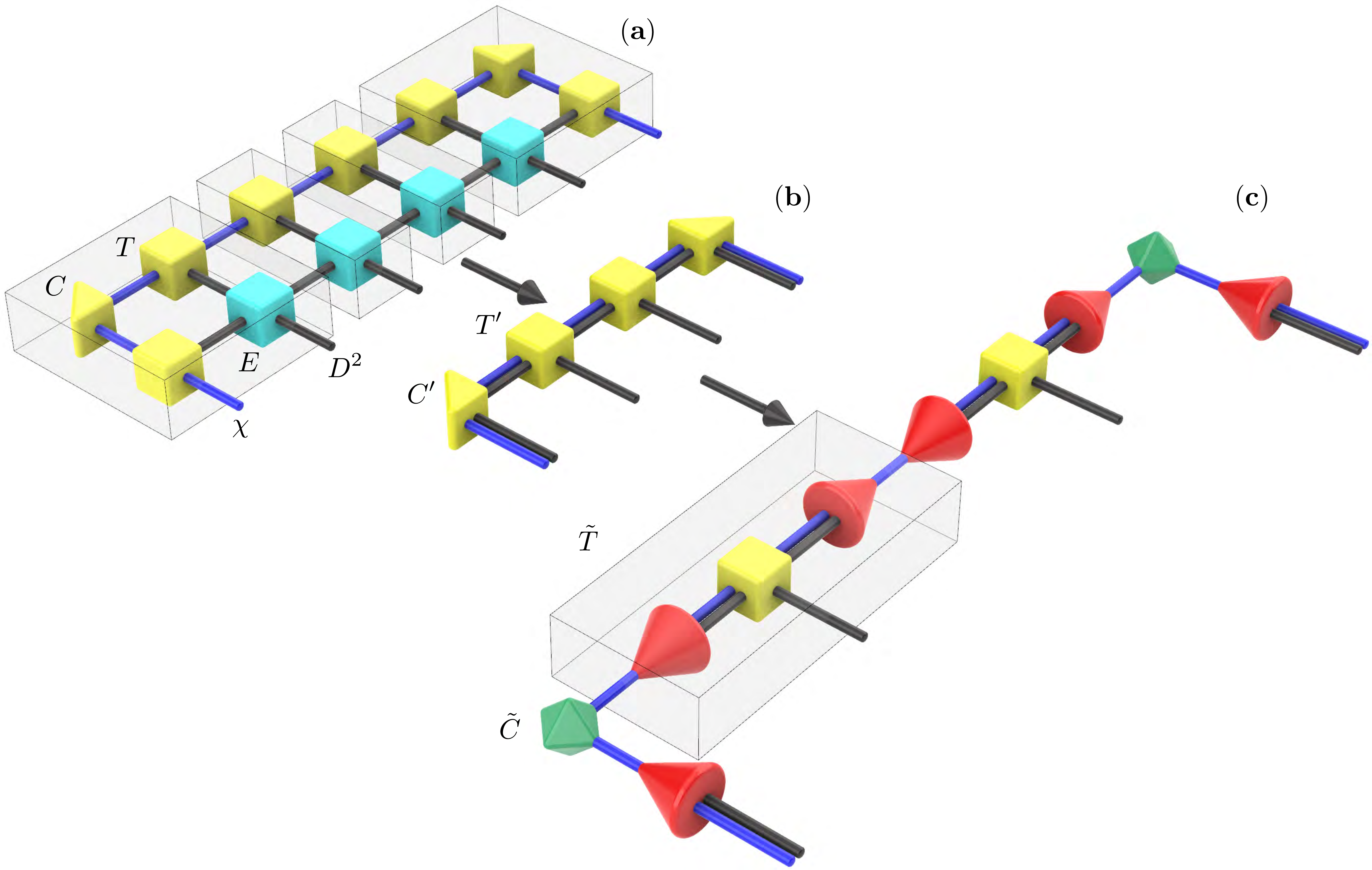}
    \caption{Single iteration of CTMRG. (a-b) It involves absorption of the single site tensor $E$ into the $C$ (corner) and $T$ (edge) tensors increasing the bond dimension from $\chi$ to $\chi D^2$ and leading to $C'$ and $E'$ tensors.  (c) Isometries (in red) obtained by orthogonal decomposition of the enlarged corner are placed on the enlarged bonds. Their absorption into the $T'$ tensor defines the renormalized $\tilde{T}$ tensor, diagonalizes the corner $C'$ to $\tilde{C}$ and reduces the dimension back to $\chi$. }
    \label{fig:CTM2}
\end{figure*}

Each renormalization step consists of three parts.
\begin{enumerate}
    \item \textit{Absorption}. At each step, the C and T tensors absorb a single site bi-layer tensor as shown in Fig.~\ref{fig:CTM2}. This raises the bond dimension of $C$ and $T$ to $\chi D^2 \times \chi D^2$ and $\chi D^2 \times \chi D^2 \times D^2$.\\
 \item  \textit{Corner renormalization}. Once absorbed, we can represent the enlarged contraction of the bi-layer network of tensors by just contracting the four identical enlarged corners. By leaving out one bond untraced, we can equate the resultant matrix to the reduced density matrix (RDM) of the system. Since this matrix is complex symmetric by construction, we can use the Orthogonal decomposition as described in Appendix A to obtain the orthogonal projectors. For stability, we just perform the orthogonal decomposition on the corner since it has the same eigenvectors as that of the RDM. We diagonalize the corner and retain the largest $\chi$ values. In order to preserve the SU(2) symmetry of the environment, we make sure that the truncation is done between values that are not part of a multiplet.\\
 \item \textit{Edge renormalization}. We use the isometries obtained from the orthogonal decomposition of the corner to renormalize the edge tensor $T$.\\
The procedure is repeated until convergence, which is identified by the measuring the change in absolute values of the corner matrix. 

\end{enumerate}

\bibliography{bibliography}

\begin{thebibliography}{35}%
\makeatletter
\providecommand \@ifxundefined [1]{%
 \@ifx{#1\undefined}
}%
\providecommand \@ifnum [1]{%
 \ifnum #1\expandafter \@firstoftwo
 \else \expandafter \@secondoftwo
 \fi
}%
\providecommand \@ifx [1]{%
 \ifx #1\expandafter \@firstoftwo
 \else \expandafter \@secondoftwo
 \fi
}%
\providecommand \natexlab [1]{#1}%
\providecommand \enquote  [1]{``#1''}%
\providecommand \bibnamefont  [1]{#1}%
\providecommand \bibfnamefont [1]{#1}%
\providecommand \citenamefont [1]{#1}%
\providecommand \href@noop [0]{\@secondoftwo}%
\providecommand \href [0]{\begingroup \@sanitize@url \@href}%
\providecommand \@href[1]{\@@startlink{#1}\@@href}%
\providecommand \@@href[1]{\endgroup#1\@@endlink}%
\providecommand \@sanitize@url [0]{\catcode `\\12\catcode `\$12\catcode
  `\&12\catcode `\#12\catcode `\^12\catcode `\_12\catcode `\%12\relax}%
\providecommand \@@startlink[1]{}%
\providecommand \@@endlink[0]{}%
\providecommand \url  [0]{\begingroup\@sanitize@url \@url }%
\providecommand \@url [1]{\endgroup\@href {#1}{\urlprefix }}%
\providecommand \urlprefix  [0]{URL }%
\providecommand \Eprint [0]{\href }%
\providecommand \doibase [0]{https://doi.org/}%
\providecommand \selectlanguage [0]{\@gobble}%
\providecommand \bibinfo  [0]{\@secondoftwo}%
\providecommand \bibfield  [0]{\@secondoftwo}%
\providecommand \translation [1]{[#1]}%
\providecommand \BibitemOpen [0]{}%
\providecommand \bibitemStop [0]{}%
\providecommand \bibitemNoStop [0]{.\EOS\space}%
\providecommand \EOS [0]{\spacefactor3000\relax}%
\providecommand \BibitemShut  [1]{\csname bibitem#1\endcsname}%
\let\auto@bib@innerbib\@empty
\bibitem [{\citenamefont {Savary}\ and\ \citenamefont
  {Balents}(2016)}]{Savary2016}%
  \BibitemOpen
  \bibfield  {author} {\bibinfo {author} {\bibfnamefont {L.}~\bibnamefont
  {Savary}}\ and\ \bibinfo {author} {\bibfnamefont {L.}~\bibnamefont
  {Balents}},\ }\bibfield  {title} {\bibinfo {title} {Quantum spin liquids: a
  review},\ }\href {https://doi.org/10.1088/0034-4885/80/1/016502} {\bibfield
  {journal} {\bibinfo  {journal} {Reports on Progress in Physics}\ }\textbf
  {\bibinfo {volume} {80}},\ \bibinfo {pages} {016502} (\bibinfo {year}
  {2016})}\BibitemShut {NoStop}%
\bibitem [{\citenamefont {Anderson}(1973)}]{Anderson1973}%
  \BibitemOpen
  \bibfield  {author} {\bibinfo {author} {\bibfnamefont {P.~W.}\ \bibnamefont
  {Anderson}},\ }\bibfield  {title} {\bibinfo {title} {Resonating valence
  bonds: A new kind of insulator?},\ }\href
  {https://doi.org/http://dx.doi.org/10.1016/0025-5408(73)90167-0} {\bibfield
  {journal} {\bibinfo  {journal} {Materials Research Bulletin}\ }\textbf
  {\bibinfo {volume} {8}},\ \bibinfo {pages} {153 } (\bibinfo {year}
  {1973})}\BibitemShut {NoStop}%
\bibitem [{\citenamefont {Poilblanc}\ \emph {et~al.}(2012)\citenamefont
  {Poilblanc}, \citenamefont {Schuch}, \citenamefont {P\'erez-Garc\'{\i}a},\
  and\ \citenamefont {Cirac}}]{Poilblanc2012}%
  \BibitemOpen
  \bibfield  {author} {\bibinfo {author} {\bibfnamefont {D.}~\bibnamefont
  {Poilblanc}}, \bibinfo {author} {\bibfnamefont {N.}~\bibnamefont {Schuch}},
  \bibinfo {author} {\bibfnamefont {D.}~\bibnamefont {P\'erez-Garc\'{\i}a}},\
  and\ \bibinfo {author} {\bibfnamefont {J.~I.}\ \bibnamefont {Cirac}},\
  }\bibfield  {title} {\bibinfo {title} {Topological and entanglement
  properties of resonating valence bond wave functions},\ }\href
  {https://doi.org/10.1103/PhysRevB.86.014404} {\bibfield  {journal} {\bibinfo
  {journal} {Phys. Rev. B}\ }\textbf {\bibinfo {volume} {86}},\ \bibinfo
  {pages} {014404} (\bibinfo {year} {2012})}\BibitemShut {NoStop}%
\bibitem [{\citenamefont {Chen}\ and\ \citenamefont
  {Poilblanc}(2018)}]{Chen2018a}%
  \BibitemOpen
  \bibfield  {author} {\bibinfo {author} {\bibfnamefont {J.-Y.}\ \bibnamefont
  {Chen}}\ and\ \bibinfo {author} {\bibfnamefont {D.}~\bibnamefont
  {Poilblanc}},\ }\bibfield  {title} {\bibinfo {title} {Topological
  $\mathbb{Z}_{2}$ resonating-valence-bond spin liquid on the square lattice},\
  }\href {https://doi.org/10.1103/PhysRevB.97.161107} {\bibfield  {journal}
  {\bibinfo  {journal} {Phys. Rev. B}\ }\textbf {\bibinfo {volume} {97}},\
  \bibinfo {pages} {161107} (\bibinfo {year} {2018})}\BibitemShut {NoStop}%
\bibitem [{\citenamefont {Rutkowski}\ and\ \citenamefont
  {Lawler}(2016)}]{Rutkowski2016}%
  \BibitemOpen
  \bibfield  {author} {\bibinfo {author} {\bibfnamefont {T.~C.}\ \bibnamefont
  {Rutkowski}}\ and\ \bibinfo {author} {\bibfnamefont {M.~J.}\ \bibnamefont
  {Lawler}},\ }\bibfield  {title} {\bibinfo {title} {Spin liquid phases of
  large-spin {M}ott insulating ultracold bosons},\ }\href
  {https://doi.org/10.1103/PhysRevB.93.094405} {\bibfield  {journal} {\bibinfo
  {journal} {Phys. Rev. B}\ }\textbf {\bibinfo {volume} {93}},\ \bibinfo
  {pages} {094405} (\bibinfo {year} {2016})}\BibitemShut {NoStop}%
\bibitem [{\citenamefont {Slagle}\ \emph {et~al.}(2022)\citenamefont {Slagle},
  \citenamefont {Liu}, \citenamefont {Aasen}, \citenamefont {Pichler},
  \citenamefont {Mong}, \citenamefont {Chen}, \citenamefont {Endres},\ and\
  \citenamefont {Alicea}}]{Alicea2022}%
  \BibitemOpen
  \bibfield  {author} {\bibinfo {author} {\bibfnamefont {K.}~\bibnamefont
  {Slagle}}, \bibinfo {author} {\bibfnamefont {Y.}~\bibnamefont {Liu}},
  \bibinfo {author} {\bibfnamefont {D.}~\bibnamefont {Aasen}}, \bibinfo
  {author} {\bibfnamefont {H.}~\bibnamefont {Pichler}}, \bibinfo {author}
  {\bibfnamefont {R.~S.~K.}\ \bibnamefont {Mong}}, \bibinfo {author}
  {\bibfnamefont {X.}~\bibnamefont {Chen}}, \bibinfo {author} {\bibfnamefont
  {M.}~\bibnamefont {Endres}},\ and\ \bibinfo {author} {\bibfnamefont
  {J.}~\bibnamefont {Alicea}},\ }\href
  {https://doi.org/10.48550/ARXIV.2204.00013} {\bibinfo {title} {Quantum spin
  liquids bootstrapped from {I}sing criticality in {R}ydberg arrays}} (\bibinfo
  {year} {2022})\BibitemShut {NoStop}%
\bibitem [{\citenamefont {Semeghini}\ \emph {et~al.}(2021)\citenamefont
  {Semeghini}, \citenamefont {Levine}, \citenamefont {Keesling}, \citenamefont
  {Ebadi}, \citenamefont {Wang}, \citenamefont {Bluvstein}, \citenamefont
  {Verresen}, \citenamefont {Pichler}, \citenamefont {Kalinowski},
  \citenamefont {Samajdar}, \citenamefont {Omran}, \citenamefont {Sachdev},
  \citenamefont {Vishwanath}, \citenamefont {Greiner}, \citenamefont
  {Vuletić},\ and\ \citenamefont {Lukin}}]{Semeghini2021}%
  \BibitemOpen
  \bibfield  {author} {\bibinfo {author} {\bibfnamefont {G.}~\bibnamefont
  {Semeghini}}, \bibinfo {author} {\bibfnamefont {H.}~\bibnamefont {Levine}},
  \bibinfo {author} {\bibfnamefont {A.}~\bibnamefont {Keesling}}, \bibinfo
  {author} {\bibfnamefont {S.}~\bibnamefont {Ebadi}}, \bibinfo {author}
  {\bibfnamefont {T.~T.}\ \bibnamefont {Wang}}, \bibinfo {author}
  {\bibfnamefont {D.}~\bibnamefont {Bluvstein}}, \bibinfo {author}
  {\bibfnamefont {R.}~\bibnamefont {Verresen}}, \bibinfo {author}
  {\bibfnamefont {H.}~\bibnamefont {Pichler}}, \bibinfo {author} {\bibfnamefont
  {M.}~\bibnamefont {Kalinowski}}, \bibinfo {author} {\bibfnamefont
  {R.}~\bibnamefont {Samajdar}}, \bibinfo {author} {\bibfnamefont
  {A.}~\bibnamefont {Omran}}, \bibinfo {author} {\bibfnamefont
  {S.}~\bibnamefont {Sachdev}}, \bibinfo {author} {\bibfnamefont
  {A.}~\bibnamefont {Vishwanath}}, \bibinfo {author} {\bibfnamefont
  {M.}~\bibnamefont {Greiner}}, \bibinfo {author} {\bibfnamefont
  {V.}~\bibnamefont {Vuletić}},\ and\ \bibinfo {author} {\bibfnamefont
  {M.~D.}\ \bibnamefont {Lukin}},\ }\bibfield  {title} {\bibinfo {title}
  {Probing topological spin liquids on a programmable quantum simulator},\
  }\href {https://doi.org/10.1126/science.abi8794} {\bibfield  {journal}
  {\bibinfo  {journal} {Science}\ }\textbf {\bibinfo {volume} {374}},\ \bibinfo
  {pages} {1242} (\bibinfo {year} {2021})},\ \Eprint
  {https://arxiv.org/abs/https://www.science.org/doi/pdf/10.1126/science.abi8794}
  {https://www.science.org/doi/pdf/10.1126/science.abi8794} \BibitemShut
  {NoStop}%
\bibitem [{\citenamefont {Giudici}\ \emph {et~al.}(2022)\citenamefont
  {Giudici}, \citenamefont {Lukin},\ and\ \citenamefont
  {Pichler}}]{giudici2022}%
  \BibitemOpen
  \bibfield  {author} {\bibinfo {author} {\bibfnamefont {G.}~\bibnamefont
  {Giudici}}, \bibinfo {author} {\bibfnamefont {M.~D.}\ \bibnamefont {Lukin}},\
  and\ \bibinfo {author} {\bibfnamefont {H.}~\bibnamefont {Pichler}},\
  }\bibfield  {title} {\bibinfo {title} {Dynamical preparation of quantum spin
  liquids in {R}ydberg atom arrays},\ }\href
  {https://doi.org/10.1103/PhysRevLett.129.090401} {\bibfield  {journal}
  {\bibinfo  {journal} {Phys. Rev. Lett.}\ }\textbf {\bibinfo {volume} {129}},\
  \bibinfo {pages} {090401} (\bibinfo {year} {2022})}\BibitemShut {NoStop}%
\bibitem [{\citenamefont {Cheng}\ \emph {et~al.}(2021)\citenamefont {Cheng},
  \citenamefont {Li},\ and\ \citenamefont {Zhai}}]{cheng2022}%
  \BibitemOpen
  \bibfield  {author} {\bibinfo {author} {\bibfnamefont {Y.}~\bibnamefont
  {Cheng}}, \bibinfo {author} {\bibfnamefont {C.}~\bibnamefont {Li}},\ and\
  \bibinfo {author} {\bibfnamefont {H.}~\bibnamefont {Zhai}},\ }\bibfield
  {title} {\bibinfo {title} {Variational approach to quantum spin liquid in a
  {R}ydberg atom simulator}\ }\href {https://doi.org/10.48550/ARXIV.2112.13688}
  {10.48550/ARXIV.2112.13688} (\bibinfo {year} {2021})\BibitemShut {NoStop}%
\bibitem [{\citenamefont {Satzinger}\ \emph {et~al.}(2021)\citenamefont
  {Satzinger}, \citenamefont {Liu}, \citenamefont {Smith}, \citenamefont
  {Knapp}, \citenamefont {Newman}, \citenamefont {Jones}, \citenamefont {Chen},
  \citenamefont {Quintana}, \citenamefont {Mi}, \citenamefont {Dunsworth},
  \citenamefont {Gidney}, \citenamefont {Aleiner}, \citenamefont {Arute},
  \citenamefont {Arya}, \citenamefont {Atalaya}, \citenamefont {Babbush},
  \citenamefont {Bardin}, \citenamefont {Barends}, \citenamefont {Basso},
  \citenamefont {Bengtsson}, \citenamefont {Bilmes}, \citenamefont {Broughton},
  \citenamefont {Buckley}, \citenamefont {Buell}, \citenamefont {Burkett},
  \citenamefont {Bushnell}, \citenamefont {Chiaro}, \citenamefont {Collins},
  \citenamefont {Courtney}, \citenamefont {Demura}, \citenamefont {Derk},
  \citenamefont {Eppens}, \citenamefont {Erickson}, \citenamefont {Faoro},
  \citenamefont {Farhi}, \citenamefont {Fowler}, \citenamefont {Foxen},
  \citenamefont {Giustina}, \citenamefont {Greene}, \citenamefont {Gross},
  \citenamefont {Harrigan}, \citenamefont {Harrington}, \citenamefont {Hilton},
  \citenamefont {Hong}, \citenamefont {Huang}, \citenamefont {Huggins},
  \citenamefont {Ioffe}, \citenamefont {Isakov}, \citenamefont {Jeffrey},
  \citenamefont {Jiang}, \citenamefont {Kafri}, \citenamefont {Kechedzhi},
  \citenamefont {Khattar}, \citenamefont {Kim}, \citenamefont {Klimov},
  \citenamefont {Korotkov}, \citenamefont {Kostritsa}, \citenamefont
  {Landhuis}, \citenamefont {Laptev}, \citenamefont {Locharla}, \citenamefont
  {Lucero}, \citenamefont {Martin}, \citenamefont {McClean}, \citenamefont
  {McEwen}, \citenamefont {Miao}, \citenamefont {Mohseni}, \citenamefont
  {Montazeri}, \citenamefont {Mruczkiewicz}, \citenamefont {Mutus},
  \citenamefont {Naaman}, \citenamefont {Neeley}, \citenamefont {Neill},
  \citenamefont {Niu}, \citenamefont {O’Brien}, \citenamefont {Opremcak},
  \citenamefont {Pató}, \citenamefont {Petukhov}, \citenamefont {Rubin},
  \citenamefont {Sank}, \citenamefont {Shvarts}, \citenamefont {Strain},
  \citenamefont {Szalay}, \citenamefont {Villalonga}, \citenamefont {White},
  \citenamefont {Yao}, \citenamefont {Yeh}, \citenamefont {Yoo}, \citenamefont
  {Zalcman}, \citenamefont {Neven}, \citenamefont {Boixo}, \citenamefont
  {Megrant}, \citenamefont {Chen}, \citenamefont {Kelly}, \citenamefont
  {Smelyanskiy}, \citenamefont {Kitaev}, \citenamefont {Knap}, \citenamefont
  {Pollmann},\ and\ \citenamefont {Roushan}}]{Satzinger2021}%
  \BibitemOpen
  \bibfield  {author} {\bibinfo {author} {\bibfnamefont {K.~J.}\ \bibnamefont
  {Satzinger}}, \bibinfo {author} {\bibfnamefont {Y.-J.}\ \bibnamefont {Liu}},
  \bibinfo {author} {\bibfnamefont {A.}~\bibnamefont {Smith}}, \bibinfo
  {author} {\bibfnamefont {C.}~\bibnamefont {Knapp}}, \bibinfo {author}
  {\bibfnamefont {M.}~\bibnamefont {Newman}}, \bibinfo {author} {\bibfnamefont
  {C.}~\bibnamefont {Jones}}, \bibinfo {author} {\bibfnamefont
  {Z.}~\bibnamefont {Chen}}, \bibinfo {author} {\bibfnamefont {C.}~\bibnamefont
  {Quintana}}, \bibinfo {author} {\bibfnamefont {X.}~\bibnamefont {Mi}},
  \bibinfo {author} {\bibfnamefont {A.}~\bibnamefont {Dunsworth}}, \bibinfo
  {author} {\bibfnamefont {C.}~\bibnamefont {Gidney}}, \bibinfo {author}
  {\bibfnamefont {I.}~\bibnamefont {Aleiner}}, \bibinfo {author} {\bibfnamefont
  {F.}~\bibnamefont {Arute}}, \bibinfo {author} {\bibfnamefont
  {K.}~\bibnamefont {Arya}}, \bibinfo {author} {\bibfnamefont {J.}~\bibnamefont
  {Atalaya}}, \bibinfo {author} {\bibfnamefont {R.}~\bibnamefont {Babbush}},
  \bibinfo {author} {\bibfnamefont {J.~C.}\ \bibnamefont {Bardin}}, \bibinfo
  {author} {\bibfnamefont {R.}~\bibnamefont {Barends}}, \bibinfo {author}
  {\bibfnamefont {J.}~\bibnamefont {Basso}}, \bibinfo {author} {\bibfnamefont
  {A.}~\bibnamefont {Bengtsson}}, \bibinfo {author} {\bibfnamefont
  {A.}~\bibnamefont {Bilmes}}, \bibinfo {author} {\bibfnamefont
  {M.}~\bibnamefont {Broughton}}, \bibinfo {author} {\bibfnamefont {B.~B.}\
  \bibnamefont {Buckley}}, \bibinfo {author} {\bibfnamefont {D.~A.}\
  \bibnamefont {Buell}}, \bibinfo {author} {\bibfnamefont {B.}~\bibnamefont
  {Burkett}}, \bibinfo {author} {\bibfnamefont {N.}~\bibnamefont {Bushnell}},
  \bibinfo {author} {\bibfnamefont {B.}~\bibnamefont {Chiaro}}, \bibinfo
  {author} {\bibfnamefont {R.}~\bibnamefont {Collins}}, \bibinfo {author}
  {\bibfnamefont {W.}~\bibnamefont {Courtney}}, \bibinfo {author}
  {\bibfnamefont {S.}~\bibnamefont {Demura}}, \bibinfo {author} {\bibfnamefont
  {A.~R.}\ \bibnamefont {Derk}}, \bibinfo {author} {\bibfnamefont
  {D.}~\bibnamefont {Eppens}}, \bibinfo {author} {\bibfnamefont
  {C.}~\bibnamefont {Erickson}}, \bibinfo {author} {\bibfnamefont
  {L.}~\bibnamefont {Faoro}}, \bibinfo {author} {\bibfnamefont
  {E.}~\bibnamefont {Farhi}}, \bibinfo {author} {\bibfnamefont {A.~G.}\
  \bibnamefont {Fowler}}, \bibinfo {author} {\bibfnamefont {B.}~\bibnamefont
  {Foxen}}, \bibinfo {author} {\bibfnamefont {M.}~\bibnamefont {Giustina}},
  \bibinfo {author} {\bibfnamefont {A.}~\bibnamefont {Greene}}, \bibinfo
  {author} {\bibfnamefont {J.~A.}\ \bibnamefont {Gross}}, \bibinfo {author}
  {\bibfnamefont {M.~P.}\ \bibnamefont {Harrigan}}, \bibinfo {author}
  {\bibfnamefont {S.~D.}\ \bibnamefont {Harrington}}, \bibinfo {author}
  {\bibfnamefont {J.}~\bibnamefont {Hilton}}, \bibinfo {author} {\bibfnamefont
  {S.}~\bibnamefont {Hong}}, \bibinfo {author} {\bibfnamefont {T.}~\bibnamefont
  {Huang}}, \bibinfo {author} {\bibfnamefont {W.~J.}\ \bibnamefont {Huggins}},
  \bibinfo {author} {\bibfnamefont {L.~B.}\ \bibnamefont {Ioffe}}, \bibinfo
  {author} {\bibfnamefont {S.~V.}\ \bibnamefont {Isakov}}, \bibinfo {author}
  {\bibfnamefont {E.}~\bibnamefont {Jeffrey}}, \bibinfo {author} {\bibfnamefont
  {Z.}~\bibnamefont {Jiang}}, \bibinfo {author} {\bibfnamefont
  {D.}~\bibnamefont {Kafri}}, \bibinfo {author} {\bibfnamefont
  {K.}~\bibnamefont {Kechedzhi}}, \bibinfo {author} {\bibfnamefont
  {T.}~\bibnamefont {Khattar}}, \bibinfo {author} {\bibfnamefont
  {S.}~\bibnamefont {Kim}}, \bibinfo {author} {\bibfnamefont {P.~V.}\
  \bibnamefont {Klimov}}, \bibinfo {author} {\bibfnamefont {A.~N.}\
  \bibnamefont {Korotkov}}, \bibinfo {author} {\bibfnamefont {F.}~\bibnamefont
  {Kostritsa}}, \bibinfo {author} {\bibfnamefont {D.}~\bibnamefont {Landhuis}},
  \bibinfo {author} {\bibfnamefont {P.}~\bibnamefont {Laptev}}, \bibinfo
  {author} {\bibfnamefont {A.}~\bibnamefont {Locharla}}, \bibinfo {author}
  {\bibfnamefont {E.}~\bibnamefont {Lucero}}, \bibinfo {author} {\bibfnamefont
  {O.}~\bibnamefont {Martin}}, \bibinfo {author} {\bibfnamefont {J.~R.}\
  \bibnamefont {McClean}}, \bibinfo {author} {\bibfnamefont {M.}~\bibnamefont
  {McEwen}}, \bibinfo {author} {\bibfnamefont {K.~C.}\ \bibnamefont {Miao}},
  \bibinfo {author} {\bibfnamefont {M.}~\bibnamefont {Mohseni}}, \bibinfo
  {author} {\bibfnamefont {S.}~\bibnamefont {Montazeri}}, \bibinfo {author}
  {\bibfnamefont {W.}~\bibnamefont {Mruczkiewicz}}, \bibinfo {author}
  {\bibfnamefont {J.}~\bibnamefont {Mutus}}, \bibinfo {author} {\bibfnamefont
  {O.}~\bibnamefont {Naaman}}, \bibinfo {author} {\bibfnamefont
  {M.}~\bibnamefont {Neeley}}, \bibinfo {author} {\bibfnamefont
  {C.}~\bibnamefont {Neill}}, \bibinfo {author} {\bibfnamefont {M.~Y.}\
  \bibnamefont {Niu}}, \bibinfo {author} {\bibfnamefont {T.~E.}\ \bibnamefont
  {O’Brien}}, \bibinfo {author} {\bibfnamefont {A.}~\bibnamefont {Opremcak}},
  \bibinfo {author} {\bibfnamefont {B.}~\bibnamefont {Pató}}, \bibinfo
  {author} {\bibfnamefont {A.}~\bibnamefont {Petukhov}}, \bibinfo {author}
  {\bibfnamefont {N.~C.}\ \bibnamefont {Rubin}}, \bibinfo {author}
  {\bibfnamefont {D.}~\bibnamefont {Sank}}, \bibinfo {author} {\bibfnamefont
  {V.}~\bibnamefont {Shvarts}}, \bibinfo {author} {\bibfnamefont
  {D.}~\bibnamefont {Strain}}, \bibinfo {author} {\bibfnamefont
  {M.}~\bibnamefont {Szalay}}, \bibinfo {author} {\bibfnamefont
  {B.}~\bibnamefont {Villalonga}}, \bibinfo {author} {\bibfnamefont {T.~C.}\
  \bibnamefont {White}}, \bibinfo {author} {\bibfnamefont {Z.}~\bibnamefont
  {Yao}}, \bibinfo {author} {\bibfnamefont {P.}~\bibnamefont {Yeh}}, \bibinfo
  {author} {\bibfnamefont {J.}~\bibnamefont {Yoo}}, \bibinfo {author}
  {\bibfnamefont {A.}~\bibnamefont {Zalcman}}, \bibinfo {author} {\bibfnamefont
  {H.}~\bibnamefont {Neven}}, \bibinfo {author} {\bibfnamefont
  {S.}~\bibnamefont {Boixo}}, \bibinfo {author} {\bibfnamefont
  {A.}~\bibnamefont {Megrant}}, \bibinfo {author} {\bibfnamefont
  {Y.}~\bibnamefont {Chen}}, \bibinfo {author} {\bibfnamefont {J.}~\bibnamefont
  {Kelly}}, \bibinfo {author} {\bibfnamefont {V.}~\bibnamefont {Smelyanskiy}},
  \bibinfo {author} {\bibfnamefont {A.}~\bibnamefont {Kitaev}}, \bibinfo
  {author} {\bibfnamefont {M.}~\bibnamefont {Knap}}, \bibinfo {author}
  {\bibfnamefont {F.}~\bibnamefont {Pollmann}},\ and\ \bibinfo {author}
  {\bibfnamefont {P.}~\bibnamefont {Roushan}},\ }\bibfield  {title} {\bibinfo
  {title} {Realizing topologically ordered states on a quantum processor},\
  }\href {https://doi.org/10.1126/science.abi8378} {\bibfield  {journal}
  {\bibinfo  {journal} {Science}\ }\textbf {\bibinfo {volume} {374}},\ \bibinfo
  {pages} {1237} (\bibinfo {year} {2021})},\ \Eprint
  {https://arxiv.org/abs/https://www.science.org/doi/pdf/10.1126/science.abi8378}
  {https://www.science.org/doi/pdf/10.1126/science.abi8378} \BibitemShut
  {NoStop}%
\bibitem [{\citenamefont {Alba}\ and\ \citenamefont
  {Calabrese}(2017)}]{Alba2017}%
  \BibitemOpen
  \bibfield  {author} {\bibinfo {author} {\bibfnamefont {V.}~\bibnamefont
  {Alba}}\ and\ \bibinfo {author} {\bibfnamefont {P.}~\bibnamefont
  {Calabrese}},\ }\bibfield  {title} {\bibinfo {title} {Entanglement and
  thermodynamics after a quantum quench in integrable systems},\ }\href
  {https://doi.org/10.1073/pnas.1703516114} {\bibfield  {journal} {\bibinfo
  {journal} {Proceedings of the National Academy of Sciences}\ }\textbf
  {\bibinfo {volume} {114}},\ \bibinfo {pages} {7947} (\bibinfo {year}
  {2017})},\ \Eprint
  {https://arxiv.org/abs/https://www.pnas.org/doi/pdf/10.1073/pnas.1703516114}
  {https://www.pnas.org/doi/pdf/10.1073/pnas.1703516114} \BibitemShut {NoStop}%
\bibitem [{\citenamefont {Robinson}\ \emph {et~al.}(2021)\citenamefont
  {Robinson}, \citenamefont {de~Klerk},\ and\ \citenamefont
  {Caux}}]{Robinson2021}%
  \BibitemOpen
  \bibfield  {author} {\bibinfo {author} {\bibfnamefont {N.~J.}\ \bibnamefont
  {Robinson}}, \bibinfo {author} {\bibfnamefont {A.~J. J.~M.}\ \bibnamefont
  {de~Klerk}},\ and\ \bibinfo {author} {\bibfnamefont {J.-S.}\ \bibnamefont
  {Caux}},\ }\bibfield  {title} {\bibinfo {title} {{On computing
  non-equilibrium dynamics following a quench}},\ }\href
  {https://doi.org/10.21468/SciPostPhys.11.6.104} {\bibfield  {journal}
  {\bibinfo  {journal} {SciPost Phys.}\ }\textbf {\bibinfo {volume} {11}},\
  \bibinfo {pages} {104} (\bibinfo {year} {2021})}\BibitemShut {NoStop}%
\bibitem [{\citenamefont {Das}(2020)}]{Das2020}%
  \BibitemOpen
  \bibfield  {author} {\bibinfo {author} {\bibfnamefont {S.~R.}\ \bibnamefont
  {Das}},\ }\href {https://doi.org/10.1093/acrefore/9780190871994.013.55}
  {\emph {\bibinfo {title} {Quantum Quench and Universal Scaling}}}\ (\bibinfo
  {publisher} {Oxford University Press},\ \bibinfo {year} {2020})\BibitemShut
  {NoStop}%
\bibitem [{\citenamefont {Mitra}(2018)}]{Aditi2018}%
  \BibitemOpen
  \bibfield  {author} {\bibinfo {author} {\bibfnamefont {A.}~\bibnamefont
  {Mitra}},\ }\bibfield  {title} {\bibinfo {title} {Quantum quench dynamics},\
  }\href {https://doi.org/10.1146/annurev-conmatphys-031016-025451} {\bibfield
  {journal} {\bibinfo  {journal} {Annual Review of Condensed Matter Physics}\
  }\textbf {\bibinfo {volume} {9}},\ \bibinfo {pages} {245} (\bibinfo {year}
  {2018})},\ \Eprint
  {https://arxiv.org/abs/https://doi.org/10.1146/annurev-conmatphys-031016-025451}
  {https://doi.org/10.1146/annurev-conmatphys-031016-025451} \BibitemShut
  {NoStop}%
\bibitem [{\citenamefont {Guardado-Sanchez}\ \emph {et~al.}(2018)\citenamefont
  {Guardado-Sanchez}, \citenamefont {Brown}, \citenamefont {Mitra},
  \citenamefont {Devakul}, \citenamefont {Huse}, \citenamefont {Schau\ss{}},\
  and\ \citenamefont {Bakr}}]{Sanchez2018}%
  \BibitemOpen
  \bibfield  {author} {\bibinfo {author} {\bibfnamefont {E.}~\bibnamefont
  {Guardado-Sanchez}}, \bibinfo {author} {\bibfnamefont {P.~T.}\ \bibnamefont
  {Brown}}, \bibinfo {author} {\bibfnamefont {D.}~\bibnamefont {Mitra}},
  \bibinfo {author} {\bibfnamefont {T.}~\bibnamefont {Devakul}}, \bibinfo
  {author} {\bibfnamefont {D.~A.}\ \bibnamefont {Huse}}, \bibinfo {author}
  {\bibfnamefont {P.}~\bibnamefont {Schau\ss{}}},\ and\ \bibinfo {author}
  {\bibfnamefont {W.~S.}\ \bibnamefont {Bakr}},\ }\bibfield  {title} {\bibinfo
  {title} {Probing the quench dynamics of antiferromagnetic correlations in a
  2d quantum {I}sing spin system},\ }\href
  {https://doi.org/10.1103/PhysRevX.8.021069} {\bibfield  {journal} {\bibinfo
  {journal} {Phys. Rev. X}\ }\textbf {\bibinfo {volume} {8}},\ \bibinfo {pages}
  {021069} (\bibinfo {year} {2018})}\BibitemShut {NoStop}%
\bibitem [{\citenamefont {Moessner}\ and\ \citenamefont
  {Sondhi}(2003)}]{Moessner2003}%
  \BibitemOpen
  \bibfield  {author} {\bibinfo {author} {\bibfnamefont {R.}~\bibnamefont
  {Moessner}}\ and\ \bibinfo {author} {\bibfnamefont {S.~L.}\ \bibnamefont
  {Sondhi}},\ }\bibfield  {title} {\bibinfo {title} {Ising and dimer models in
  two and three dimensions},\ }\href
  {https://doi.org/10.1103/PhysRevB.68.054405} {\bibfield  {journal} {\bibinfo
  {journal} {Phys. Rev. B}\ }\textbf {\bibinfo {volume} {68}},\ \bibinfo
  {pages} {054405} (\bibinfo {year} {2003})}\BibitemShut {NoStop}%
\bibitem [{\citenamefont {Moessner}\ and\ \citenamefont
  {Raman}(2011)}]{Moessner2011}%
  \BibitemOpen
  \bibfield  {author} {\bibinfo {author} {\bibfnamefont {R.}~\bibnamefont
  {Moessner}}\ and\ \bibinfo {author} {\bibfnamefont {K.~S.}\ \bibnamefont
  {Raman}},\ }\bibinfo {title} {Quantum dimer models},\ in\ \href
  {https://doi.org/10.1007/978-3-642-10589-0_17} {\emph {\bibinfo {booktitle}
  {Introduction to Frustrated Magnetism: Materials, Experiments, Theory}}},\
  \bibinfo {editor} {edited by\ \bibinfo {editor} {\bibfnamefont
  {C.}~\bibnamefont {Lacroix}}, \bibinfo {editor} {\bibfnamefont
  {P.}~\bibnamefont {Mendels}},\ and\ \bibinfo {editor} {\bibfnamefont
  {F.}~\bibnamefont {Mila}}}\ (\bibinfo  {publisher} {Springer Berlin
  Heidelberg},\ \bibinfo {address} {Berlin, Heidelberg},\ \bibinfo {year}
  {2011})\ pp.\ \bibinfo {pages} {437--479}\BibitemShut {NoStop}%
\bibitem [{\citenamefont {Mambrini}\ \emph {et~al.}(2016)\citenamefont
  {Mambrini}, \citenamefont {Or\'us},\ and\ \citenamefont
  {Poilblanc}}]{Mambrini2016}%
  \BibitemOpen
  \bibfield  {author} {\bibinfo {author} {\bibfnamefont {M.}~\bibnamefont
  {Mambrini}}, \bibinfo {author} {\bibfnamefont {R.}~\bibnamefont {Or\'us}},\
  and\ \bibinfo {author} {\bibfnamefont {D.}~\bibnamefont {Poilblanc}},\
  }\bibfield  {title} {\bibinfo {title} {Systematic construction of spin
  liquids on the square lattice from tensor networks with $\text{SU(2)}$
  symmetry},\ }\href {https://doi.org/10.1103/PhysRevB.94.205124} {\bibfield
  {journal} {\bibinfo  {journal} {Phys. Rev. B}\ }\textbf {\bibinfo {volume}
  {94}},\ \bibinfo {pages} {205124} (\bibinfo {year} {2016})}\BibitemShut
  {NoStop}%
\bibitem [{\citenamefont {Penrose}(1971)}]{Penrose1971}%
  \BibitemOpen
  \bibfield  {author} {\bibinfo {author} {\bibfnamefont {R.}~\bibnamefont
  {Penrose}},\ }\bibinfo {title} {Applications of negative dimensional
  tensors},\ in\ \href@noop {} {\emph {\bibinfo {booktitle} {Combinatorial
  Mathematics and its Applications,}}}\ (\bibinfo  {publisher} {Academic
  Press},\ \bibinfo {year} {1971})\BibitemShut {NoStop}%
\bibitem [{\citenamefont {Jordan}\ \emph {et~al.}(2008)\citenamefont {Jordan},
  \citenamefont {Or{\'{u}}s}, \citenamefont {Vidal}, \citenamefont {F.},\ and\
  \citenamefont {Cirac}}]{Jordan2008}%
  \BibitemOpen
  \bibfield  {author} {\bibinfo {author} {\bibfnamefont {J.}~\bibnamefont
  {Jordan}}, \bibinfo {author} {\bibfnamefont {R.}~\bibnamefont {Or{\'{u}}s}},
  \bibinfo {author} {\bibfnamefont {G.}~\bibnamefont {Vidal}}, \bibinfo
  {author} {\bibfnamefont {V.}~\bibnamefont {F.}},\ and\ \bibinfo {author}
  {\bibfnamefont {J.}~\bibnamefont {Cirac}},\ }\bibfield  {title} {\bibinfo
  {title} {Classical simulation of infinite-size quantum lattice systems in two
  spatial dimensions},\ }\href {https://doi.org/10.1103/PhysRevLett.101.250602}
  {\bibfield  {journal} {\bibinfo  {journal} {Phys. Rev. Letters}\ }\textbf
  {\bibinfo {volume} {101}},\ \bibinfo {pages} {250602} (\bibinfo {year}
  {2008})}\BibitemShut {NoStop}%
\bibitem [{\citenamefont {Schuch}\ \emph {et~al.}(2012)\citenamefont {Schuch},
  \citenamefont {Poilblanc}, \citenamefont {Cirac},\ and\ \citenamefont
  {P{\'{e}}rez-Garc{\'{\i}}a}}]{Schuch2012}%
  \BibitemOpen
  \bibfield  {author} {\bibinfo {author} {\bibfnamefont {N.}~\bibnamefont
  {Schuch}}, \bibinfo {author} {\bibfnamefont {D.}~\bibnamefont {Poilblanc}},
  \bibinfo {author} {\bibfnamefont {J.~I.}\ \bibnamefont {Cirac}},\ and\
  \bibinfo {author} {\bibfnamefont {D.}~\bibnamefont
  {P{\'{e}}rez-Garc{\'{\i}}a}},\ }\bibfield  {title} {\bibinfo {title}
  {{Resonating valence bond states in the PEPS formalism}},\ }\href
  {https://doi.org/10.1103/PhysRevB.86.115108} {\bibfield  {journal} {\bibinfo
  {journal} {Physical Review B}\ }\textbf {\bibinfo {volume} {86}},\ \bibinfo
  {pages} {115108} (\bibinfo {year} {2012})}\BibitemShut {NoStop}%
\bibitem [{\citenamefont {Dreyer}\ \emph {et~al.}(2020)\citenamefont {Dreyer},
  \citenamefont {Vanderstraeten}, \citenamefont {Chen}, \citenamefont
  {Verresen},\ and\ \citenamefont {Schuch}}]{Dreyer2020}%
  \BibitemOpen
  \bibfield  {author} {\bibinfo {author} {\bibfnamefont {H.}~\bibnamefont
  {Dreyer}}, \bibinfo {author} {\bibfnamefont {L.}~\bibnamefont
  {Vanderstraeten}}, \bibinfo {author} {\bibfnamefont {J.-Y.}\ \bibnamefont
  {Chen}}, \bibinfo {author} {\bibfnamefont {R.}~\bibnamefont {Verresen}},\
  and\ \bibinfo {author} {\bibfnamefont {N.}~\bibnamefont {Schuch}},\
  }\bibfield  {title} {\bibinfo {title} {Robustness of critical {U}(1) spin
  liquids and emergent symmetries in tensor networks},\ }\bibfield  {journal}
  {\bibinfo  {journal} {arXiv preprint arXiv:2008.04833}\ }\href
  {https://doi.org/10.48550/ARXIV.2008.04833} {10.48550/ARXIV.2008.04833}
  (\bibinfo {year} {2020})\BibitemShut {NoStop}%
\bibitem [{\citenamefont {Jiang}\ \emph {et~al.}(2008)\citenamefont {Jiang},
  \citenamefont {Weng},\ and\ \citenamefont {Xiang}}]{Jiang2008}%
  \BibitemOpen
  \bibfield  {author} {\bibinfo {author} {\bibfnamefont {H.~C.}\ \bibnamefont
  {Jiang}}, \bibinfo {author} {\bibfnamefont {Z.~Y.}\ \bibnamefont {Weng}},\
  and\ \bibinfo {author} {\bibfnamefont {T.}~\bibnamefont {Xiang}},\ }\bibfield
   {title} {\bibinfo {title} {Accurate determination of tensor network state of
  quantum lattice models in two dimensions},\ }\href
  {https://doi.org/10.1103/PhysRevLett.101.090603} {\bibfield  {journal}
  {\bibinfo  {journal} {Phys. Rev. Lett.}\ }\textbf {\bibinfo {volume} {101}},\
  \bibinfo {pages} {090603} (\bibinfo {year} {2008})}\BibitemShut {NoStop}%
\bibitem [{\citenamefont {Autonne}(1915)}]{Autonne1915}%
  \BibitemOpen
  \bibfield  {author} {\bibinfo {author} {\bibfnamefont {L.}~\bibnamefont
  {Autonne}},\ }\bibfield  {title} {\bibinfo {title} {Sur les matrices
  hypohermitiennes et sur les matrices unitaires},\ }\href
  {https://www.biodiversitylibrary.org/item/192858#page/9/mode/1up} {\bibfield
  {journal} {\bibinfo  {journal} {Ann. Univ. Lyon}\ }\textbf {\bibinfo {volume}
  {38}},\ \bibinfo {pages} {1} (\bibinfo {year} {1915})}\BibitemShut {NoStop}%
\bibitem [{\citenamefont {Takagi}(1924)}]{Takagi1924}%
  \BibitemOpen
  \bibfield  {author} {\bibinfo {author} {\bibfnamefont {T.}~\bibnamefont
  {Takagi}},\ }\bibfield  {title} {\bibinfo {title} {On an algebraic problem
  related to an analytic theorem of {C}arathéodory and {F}ejér and on an
  allied theorem of {L}andau},\ }\href
  {https://doi.org/doi:10.4099/jjm1924.1.0_83} {\bibfield  {journal} {\bibinfo
  {journal} {Jpn. J. Math.}\ }\textbf {\bibinfo {volume} {1}},\ \bibinfo
  {pages} {83} (\bibinfo {year} {1924})}\BibitemShut {NoStop}%
\bibitem [{\citenamefont {Suzuki}(1990)}]{SUZUKI1990}%
  \BibitemOpen
  \bibfield  {author} {\bibinfo {author} {\bibfnamefont {M.}~\bibnamefont
  {Suzuki}},\ }\bibfield  {title} {\bibinfo {title} {Fractal decomposition of
  exponential operators with applications to many-body theories and {M}onte
  {C}arlo simulations},\ }\href
  {https://doi.org/https://doi.org/10.1016/0375-9601(90)90962-N} {\bibfield
  {journal} {\bibinfo  {journal} {Physics Letters A}\ }\textbf {\bibinfo
  {volume} {146}},\ \bibinfo {pages} {319} (\bibinfo {year}
  {1990})}\BibitemShut {NoStop}%
\bibitem [{\citenamefont {Nishino}\ and\ \citenamefont
  {Okunishi}(1996)}]{Nishino1996}%
  \BibitemOpen
  \bibfield  {author} {\bibinfo {author} {\bibfnamefont {T.}~\bibnamefont
  {Nishino}}\ and\ \bibinfo {author} {\bibfnamefont {K.}~\bibnamefont
  {Okunishi}},\ }\bibfield  {title} {\bibinfo {title} {Corner transfer matrix
  renormalization group method},\ }\href {https://doi.org/10.1143/JPSJ.65.891}
  {\bibfield  {journal} {\bibinfo  {journal} {Journal of the Physical Society
  of Japan}\ }\textbf {\bibinfo {volume} {65}},\ \bibinfo {pages} {891}
  (\bibinfo {year} {1996})}\BibitemShut {NoStop}%
\bibitem [{\citenamefont {Nishino}\ \emph {et~al.}(1996)\citenamefont
  {Nishino}, \citenamefont {Okunishi},\ and\ \citenamefont
  {Kikuchi}}]{NISHINO199669}%
  \BibitemOpen
  \bibfield  {author} {\bibinfo {author} {\bibfnamefont {T.}~\bibnamefont
  {Nishino}}, \bibinfo {author} {\bibfnamefont {K.}~\bibnamefont {Okunishi}},\
  and\ \bibinfo {author} {\bibfnamefont {M.}~\bibnamefont {Kikuchi}},\
  }\bibfield  {title} {\bibinfo {title} {Numerical renormalization group at
  criticality},\ }\href
  {https://doi.org/https://doi.org/10.1016/0375-9601(96)00128-4} {\bibfield
  {journal} {\bibinfo  {journal} {Physics Letters A}\ }\textbf {\bibinfo
  {volume} {213}},\ \bibinfo {pages} {69} (\bibinfo {year} {1996})}\BibitemShut
  {NoStop}%
\bibitem [{\citenamefont {Or{\'{u}}s}\ and\ \citenamefont
  {Vidal}(2009)}]{Orus2009}%
  \BibitemOpen
  \bibfield  {author} {\bibinfo {author} {\bibfnamefont {R.}~\bibnamefont
  {Or{\'{u}}s}}\ and\ \bibinfo {author} {\bibfnamefont {G.}~\bibnamefont
  {Vidal}},\ }\bibfield  {title} {\bibinfo {title} {{Simulation of
  two-dimensional quantum systems on an infinite lattice revisited: Corner
  transfer matrix for tensor contraction}},\ }\href
  {https://doi.org/10.1103/PhysRevB.80.094403} {\bibfield  {journal} {\bibinfo
  {journal} {Physical Review B}\ }\textbf {\bibinfo {volume} {80}},\ \bibinfo
  {pages} {094403} (\bibinfo {year} {2009})}\BibitemShut {NoStop}%
\bibitem [{\citenamefont {Or{\'{u}}s}(2012)}]{Orus2012}%
  \BibitemOpen
  \bibfield  {author} {\bibinfo {author} {\bibfnamefont {R.}~\bibnamefont
  {Or{\'{u}}s}},\ }\bibfield  {title} {\bibinfo {title} {{Exploring corner
  transfer matrices and corner tensors for the classical simulation of quantum
  lattice systems}},\ }\href {https://doi.org/10.1103/PhysRevB.85.205117}
  {\bibfield  {journal} {\bibinfo  {journal} {Physical Review B}\ }\textbf
  {\bibinfo {volume} {85}},\ \bibinfo {pages} {205117} (\bibinfo {year}
  {2012})}\BibitemShut {NoStop}%
\bibitem [{\citenamefont {Calabrese}\ and\ \citenamefont
  {Cardy}(2004)}]{Calabrese_2004}%
  \BibitemOpen
  \bibfield  {author} {\bibinfo {author} {\bibfnamefont {P.}~\bibnamefont
  {Calabrese}}\ and\ \bibinfo {author} {\bibfnamefont {J.}~\bibnamefont
  {Cardy}},\ }\bibfield  {title} {\bibinfo {title} {Entanglement entropy and
  quantum field theory},\ }\href
  {https://doi.org/10.1088/1742-5468/2004/06/p06002} {\bibfield  {journal}
  {\bibinfo  {journal} {Journal of Statistical Mechanics: Theory and
  Experiment}\ }\textbf {\bibinfo {volume} {2004}},\ \bibinfo {pages} {P06002}
  (\bibinfo {year} {2004})}\BibitemShut {NoStop}%
\bibitem [{\citenamefont {Calabrese}\ \emph {et~al.}(2009)\citenamefont
  {Calabrese}, \citenamefont {Cardy},\ and\ \citenamefont
  {Doyon}}]{Calabrese2009}%
  \BibitemOpen
  \bibfield  {author} {\bibinfo {author} {\bibfnamefont {P.}~\bibnamefont
  {Calabrese}}, \bibinfo {author} {\bibfnamefont {J.}~\bibnamefont {Cardy}},\
  and\ \bibinfo {author} {\bibfnamefont {B.}~\bibnamefont {Doyon}},\ }\bibfield
   {title} {\bibinfo {title} {Entanglement entropy in extended quantum
  systems},\ }\href {http://stacks.iop.org/1751-8121/42/i=50/a=500301}
  {\bibfield  {journal} {\bibinfo  {journal} {Journal of Physics A:
  Mathematical and Theoretical}\ }\textbf {\bibinfo {volume} {42}},\ \bibinfo
  {pages} {500301} (\bibinfo {year} {2009})}\BibitemShut {NoStop}%
\bibitem [{\citenamefont {Chebotarev}\ and\ \citenamefont
  {Teretenkov}(2014)}]{CHEBOTAREV2014380}%
  \BibitemOpen
  \bibfield  {author} {\bibinfo {author} {\bibfnamefont {A.~M.}\ \bibnamefont
  {Chebotarev}}\ and\ \bibinfo {author} {\bibfnamefont {A.~E.}\ \bibnamefont
  {Teretenkov}},\ }\bibfield  {title} {\bibinfo {title} {Singular value
  decomposition for the $\text{Takagi}$ factorization of symmetric matrices},\
  }\href {https://doi.org/https://doi.org/10.1016/j.amc.2014.01.170} {\bibfield
   {journal} {\bibinfo  {journal} {Applied Mathematics and Computation}\
  }\textbf {\bibinfo {volume} {234}},\ \bibinfo {pages} {380} (\bibinfo {year}
  {2014})}\BibitemShut {NoStop}%
\bibitem [{\citenamefont {Strang}(2009)}]{strang09}%
  \BibitemOpen
  \bibfield  {author} {\bibinfo {author} {\bibfnamefont {G.}~\bibnamefont
  {Strang}},\ }\href@noop {} {\emph {\bibinfo {title} {Introduction to Linear
  Algebra}}},\ \bibinfo {edition} {4th}\ ed.\ (\bibinfo  {publisher}
  {Wellesley-Cambridge Press},\ \bibinfo {address} {Wellesley, MA},\ \bibinfo
  {year} {2009})\BibitemShut {NoStop}%
\bibitem [{\citenamefont {Eckart}\ and\ \citenamefont
  {Young}(1936)}]{Eckart1936}%
  \BibitemOpen
  \bibfield  {author} {\bibinfo {author} {\bibfnamefont {C.}~\bibnamefont
  {Eckart}}\ and\ \bibinfo {author} {\bibfnamefont {G.}~\bibnamefont {Young}},\
  }\bibfield  {title} {\bibinfo {title} {The approximation of one matrix by
  another of lower rank},\ }\href {https://doi.org/10.1007/BF02288367}
  {\bibfield  {journal} {\bibinfo  {journal} {Psychometrika}\ }\textbf
  {\bibinfo {volume} {1}},\ \bibinfo {pages} {211} (\bibinfo {year}
  {1936})}\BibitemShut {NoStop}%
\end{thebibliography}%
\end{document}